\documentclass[a4paper,11pt]{article}

\usepackage{jheppub}
\usepackage{amsmath}
\usepackage{amsfonts}
\usepackage{amssymb}

\usepackage{graphicx}% Include figure files
\usepackage{dcolumn}% Align table columns on decimal point
\usepackage{bm}% bold math
\usepackage{epsfig}

\newcommand{\dhd}{{\textstyle d}
	\lower.03ex\hbox{\kern-0.38em$^{\scriptstyle-}$}\kern-0.05em{}}
\newcommand{\dbar}{{\textstyle \delta}
	\lower.03ex\hbox{\kern-0.38em$^{\scriptstyle-}$}\kern-0.05em{}}
\newcommand{\half}{{1\over 2}}

\newcommand{\baru}{{\bar u}}

\newcommand{\bamma}{{\bar \gamma}}

\newcommand{\calf}{{\cal F}}

\newcommand{\calm}{{\cal M}}
 
\newcommand{\calo}{{\cal O}}

\newcommand{\cals}{{\cal S}} 
 
\newcommand{\calu}{{\cal U}} 
\newcommand{\calv}{{\cal V}}

\newcommand{\barpsi}{{\bar \psi}}

\newcommand{\hatp}{{\hat p}} 
\newcommand{\hatx}{{\hat x}} 
\newcommand{\haty}{{\hat y}} 
\newcommand{\hatz}{{\hat z}}

\newcommand \ket [1] {|{#1}\rangle}
\newcommand \bra [1] {\langle {#1}|}

\newcommand{\Tr}{{\rm Tr}} 
\newcommand{\tr}{{\rm tr}}

\newcommand{\ie}{i\epsilon}

\begin{document}
	
	\title{Pseudo and quasi gluon PDF in the BFKL approximation}

\author[]{Giovanni Antonio Chirilli}
\affiliation[]{Institut f\"ur Theoretische Physik, Universit\"at Regensburg,\\
	Universit\"atsstrasse 31, D-93040 Regensburg, Germany}
\emailAdd{giovanni.chirilli@ur.de}

\abstract{
I study the behavior of the gauge-invariant gluon bi-local operator with space-like separation at large longitudinal distances. 
Performing the Fourier transform,
I also calculate the behavior of the pseudo and quasi gluon PDF at low Bjorken $x$ and compare it with the leading and next-to-leading twist approximation. I show that the pseudo-PDF and quasi-PDF are very different at this regime
and that the higher twist corrections of the quasi-PDF come in not as inverse powers of $P$ but
as inverse powers of $x_B P$. 
}

\maketitle
	
\section{Introduction}

The possibility to obtain the $x_B$ dependence of the parton distribution functions (PDFs) from
lattice calculations has dragged a lot of attention in recent years 
(see \cite{Cichy:2018mum, Ji:2020ect, Cichy:2021lih} for a review). 

Since lattice gauge theory is formulated in Euclidean space, the direct calculation of the PDFs would be
impossible for objects that are defined through the light-cone matrix element of gauge-invariant bi-local operators.
For this reason, it is convenient to consider equal-time correlators and
perform the Lattice analysis in coordinate space through the Ioffe-time 
distributions~\cite{Braun:2007wv, Bali:2017gfr, Bali:2018spj, DelDebbio:2020rgv}.
Other coordinate-space based approach include, for example the ``good lattice cross sections''~\cite{Ma:2014jla, Ma:2017pxb}.
Taking the Fourier transform in momentum space, one may introduce the
quasi-PDF~\cite{Ji:2013dva} or the pseudo-PDF~\cite{Radyushkin:2017cyf, Orginos:2017kos, Joo:2020spy, HadStruc:2021wmh}.

The PDFs are extracted from the quasi-PDFs in the infinite-momentum limit $P\to\infty$ with
higher twist corrections expected to come in as inverse powers of $P$~\cite{Ji:2013dva, Ji:2020ect}.
From the pseudo-PDFs, instead, the PDFs are extracted in the short (longitudinal) distance 
limit~\cite{Radyushkin:2017cyf, Orginos:2017kos, Joo:2020spy}.

Lattice calculations provide values of the Ioffe-time distributions for a limited range of the distance 
separating the bi-local operators. 
In order to perform the Fourier transform 
for the quasi-PDF or the pseudo-PDF, it is then necessary to extrapolate the 
large-distance behavior~\cite{Braun:2007wv, Bali:2017gfr, DelDebbio:2020rgv}.

In this work the goal is to study the behavior of the Ioffe-time distribution at large longitudinal distances as well as 
the low-$x_B$ behavior of the quasi-PDF and pseudo-PDF.
To this end, we will adopt the high-energy operator product expansion (HE-OPE) formalism
(see \cite{Balitsky:2001gj} for a review) which, being formulated in coordinate space, is suitable to reach our goal.

Using the light-ray operators, obtained as analytic continuation of local twist-two operators
~\cite{BALITSKY:2014zza, Balitsky:2013npa, Balitsky:2015tca, Balitsky:2015oux, Balitsky:2018irv},
we will calculate the behavior of the leading twist (LT) and next-to-leading twist 
(NLT) contributions for the gluon Ioffe-time distribution at large longitudinal distances as well as for the
 pseudo-PDF and quasi-PDF at low-$x_B$, and
compare them with the behavior given by the BFKL resummation result. 

The paper is organized as follow.
In section \ref{sec:defoperator} we define the dimensionless gluon Ioffe-time distribution, the pseudo-PDF, and quasi-PDF.
The high-energy operator product expansion is 
reviewed in section \ref{sec:HEOPE}. In section \ref{sec:numerical} we provide the numerical values
of the parameter used throughout the paper.  In section \ref{sec:saddapproxy} we obtain the large longitudinal distance behavior of the
Ioffe-time gluon distribution, and in section \ref{sec:LT-NLT} we obtain the 
LT and NLT corrections.
The gluon pseudo-PDF and the quasi-PDF at low-$x_B$ are obtained in sections \ref{sec:pseudopdf} and \ref{sec:quasipdf} respectively.
In section \ref{sec:conclusions} we summarize our findings.

\section{Ioffe-time distribution, pseudo-PDF, and quasi-PDF}
\label{sec:defoperator}

In references~\cite{Balitsky:2019krf, Balitsky:2021qsr} the gluon matrix element with open Lorentz indexes was decomposed in terms of
six independent tensor structures built from the proton momentum $P^\mu$, the coordinate $z^\mu$ and the
antisymmetric tensor $g^{\mu\nu}$ as follow
\begin{eqnarray}
M_{\mu\alpha;\lambda\beta}\!\!\!&&\equiv \langle P|G_{\mu\alpha}(z)[z,0]G_{\lambda\beta}(0)|P\rangle
\nonumber\\
&&=I_{1\mu\alpha;\lambda\beta}{\cal M}_{pp} + I_{2\mu\alpha;\lambda\beta}{\cal M}_{zz} + I_{3\mu\alpha;\lambda\beta}{\cal M}_{zp}
\nonumber\\
&& ~~~+ I_{4\mu\alpha;\lambda\beta}{\cal M}_{pz} + I_{5\mu\alpha;\lambda\beta}{\cal M}_{ppzz} 
+ I_{6\mu\alpha;\lambda\beta}{\cal M}_{gg}\,.
\label{tensordecomp}
\end{eqnarray}
where the explicit expression of tensor structures $I_i$ with $i=1,\dots,6$ can be found in~\cite{Balitsky:2019krf}.
The amplitudes	$\cal M$ are functions of the invariants $z^2$ and $z\cdot P$.

The light-cone gluon distribution is determined from 
$g^{\alpha\beta}_\perp M_{+\alpha;+\beta}(z^+,P)$ with $z$ taken on the light-cone and proportional to the invariant 
amplitude $\calm_{pp}$
\begin{eqnarray}
g^{\alpha\beta}_\perp M_{+\alpha;+\beta}(z^+,P) = 2(P^-)^2\calm_{pp}(z\cdot p,0)\,.
\label{lightcone-pp}
\end{eqnarray}
Here we introduced the light-cone coordinate $x^\pm = {x^0\pm x^3\over \sqrt{2}}$,
and light-cone vectors $n^\mu$ and $n'^\mu$ such that $n\cdot n'=1$, $n\cdot x = x^-$ and $n'\cdot x=x^+$.

The gluon PDF $D_g(x_B)$ is then related to the amplitude $\calm_{pp}$ by
\begin{eqnarray}
\calm_{pp}(z\!\cdot\! P,0) = \half\int_{-1}^1 dx_B\, e^{iz\cdot P\, x_B}\, x_B D_g(x_B)\,.
\label{PDF-Mpp}
\end{eqnarray}
The distribution $\calm_{pp}(z\cdot P, z^2)$ is the gluon Ioffe-time distribution with $z\!\cdot\! P$ the Ioffe-time.
In section \ref{sec:saddapproxy}, we will show that the Ioffe-time plays the role of rapidity whose large logarithms are
resummed through BFKL equation.

In~\cite{Balitsky:2019krf} it was also shown that the gluon Ioffe-time distribution
is given in terms of the zeroth and transverse components as
\begin{eqnarray}
M_{0i;i0} + M_{ji;ij} = 2 P_0^2{\cal M}_{pp}\,.
\label{zero-trnsv-pp}
\end{eqnarray}

It is known that at high energy (Regge) limit
the transverse components are suppressed while the 0th and 3rd components cannot be distinguished.
Therefore, calculating the behavior of the left-hand-side (LHS) of (\ref{lightcone-pp}), will be equivalent,
at high-energy, to calculating the behavior of LHS of (\ref{zero-trnsv-pp}).

The momentum-density pseudo-PDF is defined as the Fourier transform with respect to $z\cdot P$, that is
a Fourier transform with respect to $P$ keeping its orientation fixed. So, we define the gluon pseudo-PDF as
\begin{eqnarray}
G_{\rm p}(x_B, z^2) = \int {d\varrho\over 2\pi}\, e^{-i\varrho\, x_B} \calm_{pp}(\varrho,z^2)\,,
\label{Gp}
\end{eqnarray}
where we defined $\varrho\equiv z\cdot P$.

The momentum-density quasi-PDF is defined, instead, as the Fourier transform with respect to $z^\mu$ keeping its orientation fixed. 
Let us define the vector $\xi^\mu = {z^\mu\over |z|}$, and $P_\xi = P\cdot \xi$. The quasi-PDF is then defined as
\begin{eqnarray}
G_q(x_B, P_\xi) = P_\xi\int {d\varsigma\over 2\pi}\, e^{-i\varsigma\, P_\xi x_B } \calm_{pp}(\varsigma P_\xi,\varsigma^2)\,.
\label{Gq}
\end{eqnarray}

We will calculate the large distance (large $\varrho$) behavior of the Ioffe-time distribution $\calm_{pp}(\varrho,z^2)$,
and the low-$x_B$ behavior of the gluon pseudo-PDF (\ref{Gp}), and of the gluon quasi-PDF (\ref{Gq}).

\section{High-energy OPE}
\label{sec:HEOPE}

In this section we review the HE-OPE formalism applied to the T-product of two 
electromagnetic currents in deep inelastic scattering (DIS). This provides a smooth transition
to the application of the HE-OPE to the twist-two gluon operator.

Consider the T-product of two electromagnetic currents in the background of gluon field which,
in the spectator frame, reduces to a shock wave. In deep inelastic scattering (DIS), the virtual photon, which  mediates the interactions
between the lepton and the nucleon (or nucleus), 
splits into a quark anti-quark pair long before the interaction with the target and the propagation of the dipole pair
in the shock wave background, generates two infinite Wilson lines (see \cite{Balitsky:2001gj}
for review).

\begin{figure}[htb]
	\begin{center}
		\includegraphics[width=2.0in]{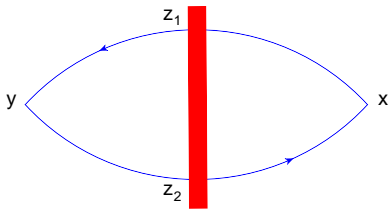}
		\caption{Diagram for the LO impact factor.
			We indicate in blue the quantum fields and in red the classical background ones.}
		\label{Fig:loif}
	\end{center}
\end{figure}

To calculate the impact factor (see Fig. \ref{Fig:loif}) we need the quark propagator in the background of gluon shock-wave.
If we consider the case $x^+>0>y^+$ the propagator is
\begin{eqnarray}
	\langle\psi(x)\barpsi(y)\rangle \stackrel{x^+>0>y^+}{=} 
	\int d^4z\delta(z_*){\hatx-\hatz\over 2\pi^2[(x-z)^2-i\epsilon]^2}
	\hatp_2U_z{\haty-\hatz\over 2\pi^2[(y-z)^2-i\epsilon]^2}\,.
	\label{LO-DIS-OPE}
\end{eqnarray}

Performing the functional integration over the spinor fields using the quark propagator (\ref{LO-DIS-OPE}),
the T-product of two electromagnetic currents
is given in terms of the impact factor, which
is related to the probability for the virtual photon to split into a quark-anti-quark pair, and matrix elements of
Wilson line which takes into accounts the recoil of the target. 
Hence, we can write
\begin{eqnarray}
\hspace{-1.3cm}&&
T\{\bar{\hat{\psi}}(x)\gamma^\mu\hat{\psi}(x)\bar{\hat{\psi}}(y)\gamma^\nu\hat{\psi}(y)\}
=\int\! d^2z_1d^2z_2~I_{\rm LO}^{\mu\nu}(z_1,z_2,x,y)
{\rm Tr}\{\hat{U}^\eta_{z_1}\hat{U}^{\dagger\eta}_{z_2}\}\,.
\label{LO-OPE}
\end{eqnarray}

To complete the HE-OPE procedure applied to DIS case, we need to find, employing the background field method, the evolution equation
of the operator ${\rm Tr}\{\hat{U}^\eta_{z_1}\hat{U}^{\dagger\eta}_{z_2}\}$ with respect to the rapidity parameter.
The evolution equation that one obtains is the non-linear BK-JIMWLK equation \cite{Balitsky:1995ub, Kovchegov:1999yj, Jalilian-Marian:1997gr, Ferreiro:2001qy, Iancu:2000hn}
\begin{eqnarray}
\hspace{-0.8cm}{d\over d\eta} \calu^\eta(x_\perp,y_\perp) =\!\!\!&& {\alpha_sN_c\over 2\pi}\!\!\int d^2z\,
{(x-y)^2_\perp\over (x-z)^2_\perp(y-z)^2_\perp}
\nonumber\\
&&\times\Big[
\calu^\eta(x_\perp,z_\perp) + \calu^\eta(y_\perp,z_\perp) - \calu^\eta(y_\perp,x_\perp) 
- \calu^\eta(x_\perp,z_\perp)\calu^\eta(y_\perp,z_\perp)
\Big]\,,
\label{BKeq}
\end{eqnarray}
where we used the notation $(x,y)_\perp = x^1y^1+x^2y^2$ and where
$\calu(x_\perp,y_\perp) = 1-{1\over N_c}\tr\{U(x_\perp)$ $U^\dagger(y_\perp)\}$.

The BFKL equation, obtained as the linearization of the non-linear equation (\ref{BKeq}) is 
\begin{eqnarray}
&&\hspace{-2cm}2a{d\over da}\,\calv_a(z_\perp) 	
= {\alpha_s N_c\over \pi^2}\!\int\!d^2z'
\Big[{\calv_a(z'_\perp)\over (z-z')_\perp^2} - {(z,z')_\perp\calv_a(z_\perp)\over z'^2_\perp(z-z')^2_\perp}\Big]\,,
\label{calvevolution}
\end{eqnarray}
where we defined 
\begin{eqnarray}
&&{1\over z^2_\perp}\calu(z_\perp) \equiv \calv(z_\perp) \,,
\label{def-calv-nu}
\end{eqnarray}
with $\calu(z_\perp)$ the forward dipole operator which depends only on its
transverse size.

Note that, evolution equation (\ref{calvevolution}) is written in terms of the parameter $a$ given by
\begin{eqnarray}
a = -{2x^+y^+\over (x-y)^2a_0}+i\epsilon\,.
\end{eqnarray}
The transition from parameter $\eta$ to parameter $a$ is detailed explained in Refs. 
\cite{Balitsky:2009xg, Balitsky:2009yp, Balitsky:2010ze, Balitsky:2012bs} where the notion of
composite conformal Wilson line operator is introduced. The motivation for this transition is to restore conformal invariance
of the NLO BK evolution equation and of the NLO impact factor.
Indeed, the evolution parameter $a$ depends on the variables $x^+$, and $y^+$ which are a remnant of the fact that
its coordinate dependence was chosen so that the NLO impact factor, which instead depends on these variable, becomes 
$SL(2,C)$ conformal invariance. The parameter $a_0$ is the initial point of the evolution.
The composite Wilson line operator is defined in such a way that
its evolution with respect to the $\eta$ parameter is zero while its evolution equation with respect to the $a$ parameter
is the LO BK equation whose linearization is the BFKL equation given in (\ref{calvevolution}).
Therefore, in principle, equation (\ref{calvevolution}) should have been written in terms of the $[\calv^a(z_{12})]_{\rm comp}$,
but its explicit expression would be relevant only at NLO, therefore, to avoid unnecessary heavy notation, we 
will keep the simple notation $\calv^a(z_{12})$, with the understanding that it is a composite operator whose NLO evolution
is conformal invariant up to running coupling contributions.

The main point is that, the conformal properties of the NLO BK equation tell us what is the proper evolution parameter in coordinate
space. This is an important point of our analysis since we are working entirely in coordinate space.

The solution of evolution equation (\ref{calvevolution}) is
\begin{eqnarray}
	\calv^a(z_{12}) = \int\!{d\nu\over 2\pi^2}(z_{12}^2)^{-\half+i\nu} \left(a\over a_0\right)^{\aleph(\gamma)\over 2}
	\int d^2\omega(\omega^2_\perp)^{-\half -i\nu}\calv^{a_0}(\omega_\perp)\,,
	\label{solution}
\end{eqnarray}
where $\aleph(\gamma)\equiv \bar{\alpha}_s\chi(\gamma)$, with $\bar{\alpha}_s = {\alpha_s N_c\over \pi}$
and $\chi(\gamma)=  2\psi(1) - \psi(\gamma) - \psi(1-\gamma)$.
The initial point of evolution is $a_0={P^-\over M_N}$ with $M_N$ the mass of the hadronic target. 

As initial condition for the evolution, one usually evaluate the dipole matrix element in the 
Golec-Biernat-W\"usthoff (GBW)~\cite{Golec-Biernat:1998js} model or the McLerran-Venugopalan (MV)~\cite{McLerran:1993ni} model.

\section{Numerical parameters}
\label{sec:numerical}

In this section we provide the numerical values of the parameters that will be used in the subsequent sections.

As initial condition for the evolution of the Ioffe-time distribution, the pseudo-PDF and quasi-PDF,
we will use the GBW model which has an $x_B$ dependence
of the saturation scale as $Q_s = \left(x_0/ x_B\right)^{0.277\over 2}$,
with $x_0 = 0.41\times 10^{-4}$~~\cite{Golec-Biernat:1998js}. The dimension of the dipole cross-section
is given by the parameter $\sigma_0 = 29.12$\,mb~\cite{Golec-Biernat:1998js}. Our starting point for the
evolution in $x_B$ will be $x_B = 0.1$ for which $Q_s = 0.34$\,GeV. 
As we will see, $x_B=0.1$ is the point at which the BFKL logarithms start to be of order $1$ thus need to be
resummed. However, the conclusions we will reach through our analysis will not depend on the starting point of the evolution.

The value of the running coupling we choose is
 $\bar{\alpha}_s={\alpha_sN_c\over \pi}=0.2$. 
Since at low-$x_B$ the strong coupling runs with the transverse 
size of the dipole~\cite{Balitsky:2006wa, Kovchegov:2006vj, Balitsky:2007feb}, our choice of the running coupling
corresponds to a transverse size of about $0.2\,{\rm GeV}^{-1}$ so that the running-coupling is evaluated at a momentum scale
greater than our choice of $Q_s$.

As we will show in subsequent sections, there is a correspondence between the Ioffe-time $z\cdot P$ and the $x_B$
of the type $z\cdot P = x_B^{-1}$. So, to calculate the large longitudinal distance behavior 
of the Ioffe-time distribution, we use as starting point of the 
evolution (in the Ioffe-time parameter) the value $\varrho\equiv z\cdot P = 10$ and we fix $|z|= 0.5\, {\rm GeV}^{-1}$.

For the quasi-PDF, we introduce the unit vector $\xi^\mu$ in the direction of the gauge link
and choose the projection of the target momentum $P$ along  $\xi^\mu$
to be $\xi\cdot P \equiv P_\xi = 4$\,GeV (see section \ref{sec:quasipdf} for details).

\section{Ioffe-time gluon distribution at large longitudinal distances}
\label{sec:saddapproxy}

In this section we are going to calculate the behavior of Ioffe-time gluon distribution at 
large longitudinal distances in the saddle point approximation.
We will show that large distance, \textit{i.e.} large values of the Ioffe-time $z\cdot P=\varrho$, generates large logarithms
which are resummed by the BFKL equation through the HE-OPE formalism. Therefore, we will refer to high-energy 
or large-distance behavior interchangeably.

As discussed in the previous section, the Ioffe-time gluon distribution is defined by
\begin{eqnarray}
\bra{P} {G^a}^{i-}(z)[z,0]{G^b_i}^-(0)\ket{P} = 2(P^-)^2\calm_{pp}(\varrho,z^2)\,.
\label{baliradymine}
\end{eqnarray}
In the high-energy (Regge) limit forward matrix elements of $U_xU^\dagger_y$ operator are divergent because they contain an
unrestricted integration along the $n^\mu$ direction, so we have
\begin{eqnarray}
&&\hspace{-1cm}
2\pi\delta(\epsilon^-)\bra{P} G^{a\,i-}(x^+,x_\perp)
[n x^+ + x_\perp, 0]^{ab}G_i^{b\,-}(0)\ket{P}
\nonumber\\
&&\hspace{-1cm}
~~= \int_{-\infty}^{+\infty}\!dy^+
\bra{P} G^{a\,i-}(x^++y^+,x_\perp)
\nonumber\\
&&\hspace{0cm}\times[n (x^+ +y^+) + x_\perp, ny^+ ]^{ab}G_i^{b\,-}(y^+)
\ket{P+\epsilon^-n}\,.
\label{unresctrictedint}
\end{eqnarray}

\begin{figure}[htb]
	\begin{center}
		\includegraphics[width= 2.70in]{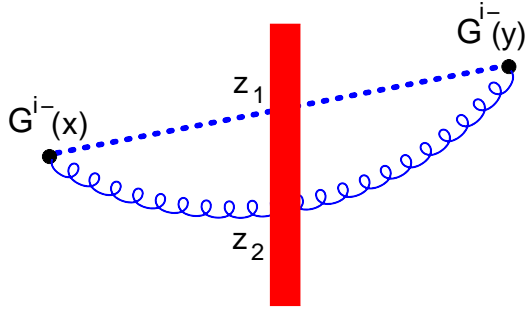}
		\caption{Diagrams for the impact factors for the quasi-PDFs operators. The black bullets represent the
			gluons and quark fields respectively.}
		\label{Fig:quasipdf-gq-LO}
	\end{center}
\end{figure}

As described in section \ref{sec:HEOPE}, the first step is 
to calculate the impact factor from the operator (\ref{unresctrictedint}).
To this end, we split all fields in quantum and classical, functionally integrate over the quantum field and obtain the diagram in figure
\ref{Fig:quasipdf-gq-LO}.

Assuming that the shock-wave is in $x^+=0$ position, we distinguish the cases $x^+>0>y^+$ and $x^+<0<y^+$.
The high-energy gluon propagator at $x^+>0>y^+$  in coordinate space~\cite{Balitsky:2009yp} is
\begin{eqnarray}
&&\hspace{-0.7cm}\langle {\rm T}A_\mu^a(x)A^b_\nu(y)\rangle 
\stackrel{x^+>0>y^+}{=} 
\nonumber\\
&& - {1\over 4\pi^3}\int d^2z\,U^{ab}_z
{x^+|y^+|g^\perp_{\mu\nu} - |y^+|n'_{\mu}X^\perp_\nu  + x^+n'_{\nu}Y^\perp_\mu + n'_{\mu}n'_{\nu}  (X,Y) \over 
\Big[2(x-y)^-x^+|y^+|- |y^+|(x-z)^2_\perp - x^+(y-z)^2_\perp + \ie\Big]^2}\,,
\label{gluonprop-coord}
\end{eqnarray}
where we defined $X_{2\perp} = (x-z_2)_\perp$ and $Y_{2\perp} = (y-z_2)_\perp$.
So, using propagator (\ref{gluonprop-coord}) diagram \ref{Fig:quasipdf-gq-LO} is
\begin{eqnarray}
	&&\hspace{-2.5cm}\langle G^{a\,i-}(x^+,x_\perp)[n x^+ + x_\perp, ny^+ + y_\perp]^{ab}G^{b\,j-}(y^+,y_\perp)
	\rangle_{{\rm Fig.}\ref{Fig:quasipdf-gq-LO}}
	\nonumber\\
	&&\hspace{-2cm}
	\stackrel{y^+<0<x^+}{=}  (\partial_x^i g^{\mu -} - \partial_x^-g^{i\mu})(\partial_y^j g^{\nu -} - \partial_y^-g^{j\nu})
	\langle A^a_\mu(x)A^b_\nu(y)\rangle_A
	\nonumber\\
	&&\hspace{-1.4cm}= - \!\int {d^2z_2\over 4\pi^3}\,U^{ab}_{z_2}U^{ab}_{z_1}\,
	(\partial_x^i g^{\mu -} - \partial_x^-g^{i\mu})(\partial_y^j g^{\nu -} - \partial_y^-g^{j\nu})
	\nonumber\\
	&&\hspace{-1.6cm}~~~\times{x^+|y^+|g^\perp_{\mu\nu} - |y^+|n'_{\mu}X^\perp_{2\nu} 
		+ x^+n'_{\nu}Y^\perp_{2\mu} + n'_{\mu}n'_{\nu}  (X_2,Y_2)\over 
		\Big[|y^+|(x-z_2)^2_\perp + x^+(y-z_2)^2_\perp - \ie\Big]^2}\,.
	\label{result-diagram}
\end{eqnarray}
Result (\ref{result-diagram}) is proportional to two Wilson lines in the adjoint representation, 
one at point $z_{2\perp}$ as a result of the gluon propagation in the shock-wave external field,
and the other one at point $z_{1\perp}$ which is the point at which the shock-wave intersects the gauge link
$[n x^+ + x_\perp, ny^+ + y_\perp]^{ab}$. The point $z_1$ can actually be anywhere between point $nx^++x_\perp$ and $ny^++y_\perp$,
so we parametrize the straight line between these two points as 
$x_u = ux_\perp + \baru y_\perp = z_{1\perp}$, with $u = {|y^+|\over \Delta^+}$, $\baru = {x^+\over \Delta^+}$
and $\Delta^+ = x^+ - y^+$, and
at the end we will have to integrate over the parameter $u$. However, for technical reason, as it will be clear later, we
will continue to call such point as $z_{1\perp}$.

After differentiation, eq. (\ref{result-diagram}) becomes
\begin{eqnarray}
	&&\hspace{-1.5cm}\langle G^{a\,i-}(x^+,x_\perp)[n x^+ + x_\perp, ny^+ + y_\perp]^{ab}G_i^{b\,-}(y^+,y_\perp)
	\rangle_{{\rm Fig.}\ref{Fig:quasipdf-gq-LO}}
	\nonumber\\
	&&\hspace{-1.5cm}~~
	\stackrel{y^+<0<x^+}{=} \!\!
	  \int {d^2z_2\over 4\pi^3}\Bigg[
	 12\,{-2 x^+y^+(x-z_2,y-z_2)^2 + x^+y^+(x-z_2)^2_\perp(y-z_2)^2_\perp\over\big [x^+(y-z_2)^2_\perp-y^+(x-z_2)^2_\perp\big]^4}
	 \Bigg]U^{ab}_{z_2}U^{ab}_{z_1} \,,
\label{dresult-diagram-differentiated}
\end{eqnarray}
As done for the photon impact factor in section \ref{sec:HEOPE}, we include the case $\theta(x^+y^+)$
 and consider forward matrix elements. Thus, we make the substitution $U^{ab}_{z_2}U^{ab}_{z_1}$ with $- 2 N^2_c\calu(z_{12})$ where
$\calu(z_{12})=1-{1\over N_c}\tr\{U_{z_{1\perp}}U^\dagger_{z_{2\perp}}\}$ and $\tr$ trace in the fundamental representation.
 
Let us observe that we have put the starting operator in the left-hand-side of eq. (\ref{result-diagram}) 
in the form of impact factor convoluted with matrix element of Wilson lines as it was in the case of the T-product of
two electromagnetic current in eq. (\ref{LO-DIS-OPE}). The last step of the high-energy OPE is to convolute the impact factor 
with the solution of the evolution equation of the Wilson-line operators, eq. (\ref{solution}). So, we have
\begin{eqnarray}
	&&\hspace{-0.6cm}g_{ij}\langle G^{a\,i-}(x^+,x_\perp)[n x^+ + x_\perp, ny^+ + y_\perp]^{ab}G^{b\,j-}(y^+,y_\perp)
	\rangle_{{\rm Fig.}\ref{Fig:quasipdf-gq-LO}}
	\nonumber\\
	&&\hspace{-0.6cm}~~
	\stackrel{y^+<0<x^+}{=} - {6N^2_c\over (x^+|y^+|)^3}
	\int{d\nu\over  2\pi^2}
	\,\left( - {2{\Delta^+}^2 u\baru\over \Delta^2_\perp }{{P^-}^2\over M^2_N} + i\epsilon \right)^{{\alpha_s N_c\over 2\pi}\chi(\nu)}
	\int d^2\omega (\omega^2_\perp)^{-\half-i\nu}\langle \calv^{a_0}(\omega_\perp)\rangle
	\nonumber\\
		&&\hspace{-0.6cm}~~~~~~
	\times\!\int {d^2z_2\over \pi^3}\,
	{-2(x-z_2,y-z_2)^2_\perp + (x-z_2)^2_\perp(y-z_2)^2_\perp
		\over \left( {(y-z_2)^2_\perp\over |y^+|} + {(x-z_2)^2_\perp\over x^+}\right)^4}(z_{12}^2)^{\half+i\nu}\,.
	\label{contractedindex}
\end{eqnarray}
In section \ref{sec:projection} we show that the matrix element
in eq. (\ref{contractedindex}) with transverse indexes $i$ and $j$ not contracted,
after projection on the power like eigenfunction $(z_{12}^2)^{\half+i\nu}$, has only one tensor structure, that is $g^{ij}$.

Here, we proceed with contracted $i$ and $j$ indexes, and obtain 
\begin{eqnarray}
&&\hspace{-2cm}\int {d^2z_2\over \pi^3}\,
{-2(x-z_2,y-z_2)^2_\perp + (x-z_2)^2_\perp(y-z_2)^2_\perp
\over \left( {(y-z_2)^2_\perp\over |y^+|} + {(x-z_2)^2_\perp\over x^+}\right)^4}
(z_{12}^2)^{\half+i\nu}
\nonumber\\
&&\hspace{-2cm}
= - {\gamma^2\over \pi^2}(\Delta^+)^4 (\Delta^2_\perp)^{\gamma-1}
\Gamma(1-\gamma)\Gamma(1+\gamma)(u\baru)^{\gamma+3} \,.
\label{final-contractedindex-projection}
\end{eqnarray}
Using (\ref{final-contractedindex-projection}) in (\ref{contractedindex}) and integrating over $u$ we arrive at
\begin{eqnarray}
	&&\hspace{-0.7cm}
	\int_{-\infty}^{+\infty}\!\!\!dx^+dy^+\!\,\delta(x^+-y^+ - L)
	\langle G^{a\,i-}(x^+,x_\perp)[n x^+ + x_\perp, ny^+ + y_\perp]^{ab}G_i^{b\,-}(y^+,y_\perp)
	\rangle_{{\rm Fig.}\ref{Fig:quasipdf-gq-LO}}
	\nonumber\\
	&&\hspace{-0.7cm}
	= 6N^2_c\int{d\nu\over  2\pi^2}
	\left(- {2 u\baru\over \Delta^2_\perp} {(LP^-)^2\over M^2_N} + i\epsilon\right)^{{\alpha_s N_c\over 2\pi}\chi(\nu)}
	\int d^2\omega (\omega^2_\perp)^{-\half-i\nu}\langle \calv^{a_0}(\omega_\perp)\rangle
	\nonumber\\
	&&\hspace{-0.7cm}~~~
	\times\!\!\int_0^1\!{du\over L}
	{\gamma^2\over  \pi^2} (\Delta^2_\perp)^{\gamma-1}
	\Gamma(1-\gamma)\Gamma(1+\gamma)(u\baru)^{\gamma}
	\nonumber\\
	&&\hspace{-0.7cm}
	= 3 N^2_c\int\!d\nu\left(-{2(LP^-)^2\over \Delta^2_\perp M^2_N}+i\epsilon\right)^{{\aleph(\gamma)\over 2}}
	{ (\Delta^2_\perp)^{\gamma-1}\over L\,\pi^4} 
	{\pi\gamma\over \sin \pi\gamma}
	{\gamma^4\Gamma^2(\gamma)\over \Gamma(2+2\gamma)} 
	\nonumber\\
	&&\hspace{-0.7cm}~~~
\times\!\!\int d^2\omega (\omega^2_\perp)^{-\half-i\nu}\langle \calv^{a_0}_{\omega}\rangle + \calo(\alpha_s)\,.
	\label{projection}
\end{eqnarray}

\begin{figure}
	\begin{center}
		\includegraphics[width=3.7in]{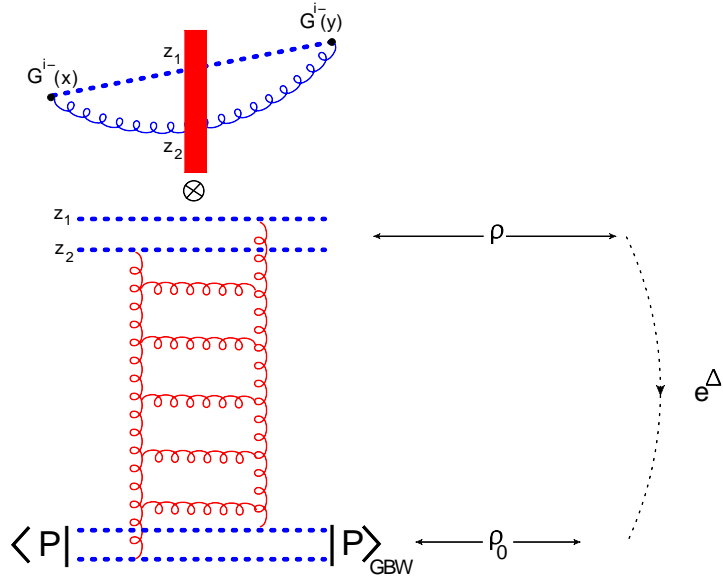}
	\end{center}
\caption{Diagrammatic representation of the HE-OPE applied to the gluon non-local operator with
``quasi-PDF frame''.}
\label{Fig:gluon-qPDFevolution}
\end{figure}

We will evaluate the integration over the parameter $\nu$ numerically and compare it with the saddle point approximation. 
The GBW model~\cite{Golec-Biernat:1998js}  applied to the Wilson-line dipole matrix element evaluated in the target nucleon state is
(see figure \ref{Fig:gluon-qPDFevolution}).
\begin{eqnarray}
\bra{P}\calu(x-y)\ket{P+\epsilon^- n'} =&& P^-2\pi\delta(\epsilon^-)\langle\bra{P}\calu(x-y)_\perp\ket{P}\rangle
\nonumber\\
=&& P^-2\pi\delta(\epsilon^-)\sigma_0\left(1 - \exp\Big({-(x-y)^2_\perp Q^2_s\over 4}\Big)\right)\,,
\label{GBW}
\end{eqnarray}
where, as discussed in section \ref{sec:numerical}, the saturation scale assume the fixed value of $Q_s = 0.34$\,GeV.

So, using (\ref{GBW}) in eq. (\ref{projection}) we arrive, with the help of eq. (\ref{unresctrictedint}), at
\begin{eqnarray}
	&&\hspace{-1cm}
	{1\over 2(P^-)^2}\bra{P} G^{a\,i-}(L,x_\perp)[n L + x_\perp, 0]^{ab}G_i^{b\,-}(0)\ket{P}
	\nonumber\\
	&&\hspace{-1cm}
	= {3N^2_c\over 2\pi^3} 
	{\sigma_0\over \Delta^2_\perp}{1\over LP^-}
	\!\int\!d\nu\left(-{2(LP^-)^2\over\Delta^2_\perp M^2_N}+i\epsilon\right)^{\aleph(\gamma)\over 2}
{\gamma\,\Gamma^2(1-\gamma)\Gamma^3(1+\gamma)\over  \Gamma(2+2\gamma)} 
	\left({Q^2_s\Delta^2_\perp\over 4}\right)^{\gamma}\,.
	\label{withmodel0}
\end{eqnarray}

Let $z^\mu$ be a space-like vector.
In the high-energy limit, the coordinate $x^+$ component is enhanced, the $x^-$ one is suppressed, 
and the $x_\perp$ component is left invariant, so
$|\Delta_\perp| = \sqrt{-z^2} = |z|>0$ with $z^2<0$.
Since at this regime we do not distinguish between the $0$ and the $3$-component,
we also have $(P^-)^2 = (P\cdot {z\over |z|})^2 = {\varrho^2\over -z^2}$
and $LP^-\to z\cdot P = \varrho $. Using variables $\varrho$ and $z^2$, and eq. (\ref{lightcone-pp}) 
with $z^2\neq 0$, result (\ref{withmodel0}) becomes
\begin{eqnarray}
\hspace{-0.7cm}\calm_{pp}(\varrho,z^2) = 
{3N^2_c\over 4\pi^3} {Q_s\sigma_0\over |z|\varrho} \!\int\!d\nu\!
\left({2\,\varrho^2\over z^2 M^2_N}+i\epsilon\right)^{\aleph(\gamma)\over 2}
{\gamma\,\Gamma^2(1-\gamma)\Gamma^3(1+\gamma)\over \Gamma(2+2\gamma)} 
\left({|z|Q_s\over 2}\right)^{\!2i\nu}\,.
\label{withmodel}
\end{eqnarray}

We are now ready to perform last integration using the saddle-point approximation technique. We
notice that
${\gamma\Gamma^2(1-\gamma)\Gamma^3(1+\gamma)\over \Gamma(2+2\gamma)}$, is a very slowly varying function, so
we have
\begin{eqnarray}
\calm_{pp}(\varrho,z^2) \simeq {3N^2_c\over 128\,\varrho}{Q_s\,\sigma_0\over |z|}
\left({2\,\varrho^2\over z^2M^2_N}+i\epsilon\right)^{\bar{\alpha}_s2\ln 2}
{e^{-{\ln^2 {Q_s|z|\over 2}\over 7\zeta(3)\bar{\alpha}_s\ln\left({2\,\varrho^2\over z^2M^2_N}+i\epsilon\right)}}
	\over \sqrt{7\zeta(3)\bar{\alpha}_s\ln\left({2\,\varrho^2\over z^2M^2_N}+i\epsilon\right)}}\,.
\label{Mpp}
\end{eqnarray}
\begin{figure}[htb]
	\begin{center}
		\includegraphics[width = 2.8in]{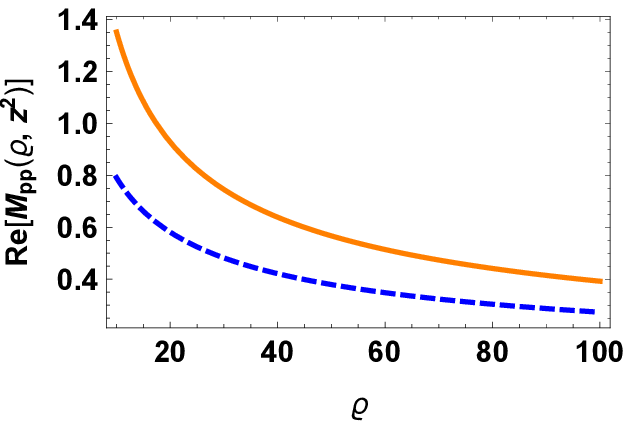}
		\includegraphics[width = 2.8in]{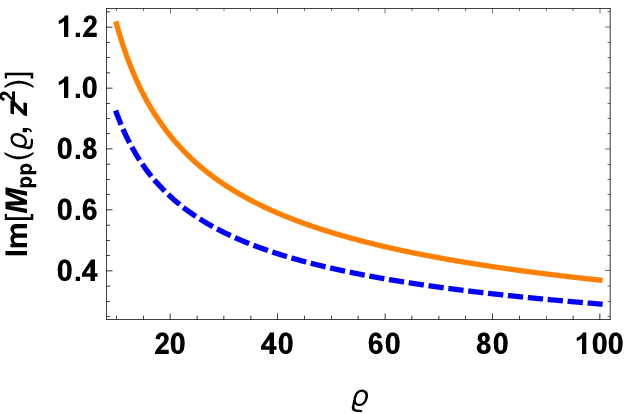}
	\end{center}
	\caption{In the left and in the right panel we plot the numerical evaluation of real and imaginary part of (\ref{withmodel}) (orange curve), 
		and the real and imaginary part of the saddle point result (\ref{Mpp}) (blue dashed curve), respectively. }
	\label{Fig:gluon-bfkl-sadVSnum}
\end{figure}

In figure \ref{Fig:gluon-bfkl-sadVSnum} we 
compare the saddle-point result (\ref{Mpp}) with the numerical integration (\ref{withmodel}).

The first thing to notice is that, the logarithms resummed by BFKL are $\bar{\alpha}_s\ln\left({\sqrt{2}\varrho\over |z|M_M}\right)$ which,
given that we are using $\bar{\alpha}_s=0.2$, $|z|=0.5\,{\rm GeV}^{-1}$, and $M_N=1$\,GeV, is of order 1 starting from $\varrho \sim 10$. We see that $\varrho$ acts like a
rapidity parameter. We evolve the distribution with $\varrho$ until $\bar{\alpha}_s\ln\left({\sqrt{2}\varrho\over |z|M_M}\right)$ is
of order $1$, that is we start with large values of $\varrho$ and end at smaller ones. The dipole at the smallest value of
$\varrho$ is evaluated in the GBW model. In our case, as anticipated in section \ref{sec:numerical}, we stop the evolution 
at $\varrho=10$ (see figure \ref{Fig:gluon-qPDFevolution}).

In Fig. \ref{Fig:gluon-bfkl-sadVSnum} we plotted the real and imaginary part of (\ref{withmodel}) and (\ref{Mpp}) 
with $\varrho\in [10,400]$. For the subsequent analysis we will use the numerical evaluation
of (\ref{withmodel}) (orange curve).

\section{Leading and Next-to-leading Twist}
\label{sec:LT-NLT}

As described in section \ref{sec:HEOPE}, within the high-energy OPE, the DIS cross section can be written as a convolution of the
impact factor and the solution of the evolution equation of the matrix elements of the dipole-Wilson-line operator which, in the 
linear case, is the BFKL equation. Therefore, the dipole DIS cross-section can be written as
\begin{eqnarray}
\sigma^{\gamma^*p}(x_B,Q^2) = \int d\nu \, F(\nu)\,x_B^{-\aleph(\nu) -1}\left({Q^2\over P^2}\right)^{\half+i\nu}\,,
\label{dipolecrossection}
\end{eqnarray}
where $\aleph(\gamma)$ is the BFKL pomeron intercept, $F(\nu)$ the pomeron residue,
and in the limit under consideration, $-q^2=Q^2\gg P^2$, and $s = (P+q)^2\gg Q^2$.

To get the dipole cross-section given in eq. (\ref{dipolecrossection}) we can calculate the integral over the $\nu$-parameter
numerically, or with the saddle point approximation capturing in this way the full BFKL dynamics. In the previous section, we have done this
for the gluon Ioffe-time distribution (see eqs. (\ref{withmodel}) and (\ref{Mpp})).

The $n$-th moment of the structure function is
\begin{eqnarray}
M_n = \int_0^1 dx_B \, x_B^{n-1} \sigma^{\gamma^*p}(x_B,Q^2)
= \int_{\half -i\infty}^{\half + i\infty} d\gamma\,{F(\gamma)\over n-1-\aleph(\gamma)}
\left({Q^2\over P^2}\right)^\gamma\,,
\nonumber
\end{eqnarray}
where $\gamma=\half + i\nu$. The integration over the $\gamma$-parameter can be performed closing
the contour to the left of the poles. In this way we get the
anomalous dimensions of the leading and higher twist operators at the \textit{unphysical point}.

In our case, to get the LT and NLT contributions from the infinite series of twists, resummed by BFKL eq., 
we could repeat the same steps outlined above.
However, this time we need the explicit form of the gluon twist-two operator at the 
unphysical point $n=1$, $F^a_{\mu_+}\nabla^{-1}_+F^{\mu a}_+$.

In Ref. \cite{BALITSKY:2014zza} (see also \cite{Balitsky:2013npa, Balitsky:2018irv}) 
it was shown that the analytical continuation of anomalous dimension of twist-two gluon operator
$\calo^j_F\equiv F_\mu^{a\,-}\nabla_+^{j-2}F^{a\,\mu-}$ to the \textit{unphysical point} 
$j=1$, which is determined by BFKL equation,
can be obtained by the analytic continuation of the operator itself at such \textit{unphysical point}.
It is known that the anomalous dimensions are singular at $j=1$, and this means that in this
limit there is a different hierarchy of perturbation theory which requires a new resummation 
of terms like $\left({\alpha_s\over j-1}\right)^n$ (at leading-log),
and therefore it implies the existence of a different (than DGLAP) evolution equation, the BFKL equation, which resums such logarithms.
For this reason we will refer to the Regge limit also as the $j\to 1$ limit.
The relation between DGLAP and BFKL equation at the unphysical point
was established, at the level of anomalous dimension, at LO in Ref \cite{Jaroszewicz:1982gr} and at NLO in Ref. \cite{Fadin:1998py}. 

In Ref. \cite{Balitsky:1987bk} it was shown that the local Operator Product Expansion can be reformulated in terms of non-local
operators with the advantage of preserving explicitly the Lorentz and the conformal invariance of the theory and
also providing a gauge covariant technique to separating higher twist contribution.
Such non-local operator are light-ray operator with integration over the longitudinal direction. In light of this,
in Ref. \cite{BALITSKY:2014zza}, it is shown how to construct analytic continuation of non-local operator at $j=1$. 
Following the procedure described in Ref. \cite{Balitsky:2018irv}, in section \ref{sec:analyticLR},
we will show that such analytic continuation of local operator is
\begin{eqnarray}
\hspace{-0.6cm}F_{p_1\xi}^a(x)\nabla^{j-2}F_{p_1}^{a~\xi}(x)\Big|_{x=0} 
\stackrel{\rm forw.}{=}  {1\over \Gamma(2-j)}\int^{\infty}_0 dv\, v^{1-j}\, F^a_{p_1\xi}(0)[0,vp_1]^{ab}F^{b~\xi}_{p_1}(vp_1)\,,
\label{integratingHminus-forw-bb}
\end{eqnarray} 
where $p_1$ is a light-cone vector and where the notation $x_{p_1} = p_1^\mu x_\mu$ has been used
(see section \ref{sec:analyticLR} for the details). 
Equation (\ref{integratingHminus-forw-bb}) is the gluon light-ray operator with spin $j$. 

It turns out that  in the high-energy (Regge) limit
the correlation function of two gluon light-ray operators (\ref{integratingHminus-forw-bb}), with spin $j$ and $j'$ respectively,
are UV divergent if taken on the light-cone. In Refs.~\cite{Balitsky:2013npa, Balitsky:2015tca, Balitsky:2015oux, Balitsky:2018irv} 
it was shown that a way to regularize this UV divergence is to consider the point-splitting regulator, that is, the 
light-ray operator, which lays on the light-cone,  becomes a rectangular frame, the ``Wilson-frame'' (see figure \ref{frames}).
In section \ref{sec:analyticLR} we will show that an alternative regulator is the
``quasi-pdf frame'' (see figure \ref{frames}) defined as
\begin{eqnarray}
&&\calf^j(x,y) = \int_0^{+\infty}du^{1-j}\,{F^{a\,i}}_{p_1}(up_1 + x)[up_1+x,y]^{ab}{F_i^{b\,}}_{p_1}(y)\,.
\label{qpdf-frame}
\end{eqnarray}

As a consistency check, we will also show that,
in the high-energy regime correlation functions of two such operators, eq. (\ref{qpdf-frame}), agrees with
the expected result from conformal field theory.

The Mellin transform of eq. (\ref{withmodel0}), which is, as explained above, the analytic 
continuation to non integer $j$ of local operator with ``quasi-pdf frame'' is 
\begin{eqnarray}
&&\int_{\Delta^2_\perp M_N}^{+\infty}\!dL \, L^{1-j}\int_{-\infty}^{+\infty}\!dy^+
\bra{P} G^{a\,i-}(L+y^+,x_\perp)
\nonumber\\
&&
~~~~~~~~~~~~~~~~~~~~~~~~~
\times[n (L +y^+) + x_\perp, ny^+ + y_\perp]^{ab}G_i^{b\,-}(y^+,y_\perp)\ket{P}
\nonumber\\
&&
=-{3\,i\,N^2_c\sigma_0P^-\over \pi^3\Delta^2_\perp}\int_{\half-i\infty}^{\half+i\infty}\!d\gamma \,{\theta(\Re[j-1-\aleph(\gamma)])\over \omega - \aleph(\gamma)} 
{\gamma\Gamma^2(1-\gamma)\Gamma^3(1+\gamma)\over \Gamma(2+2\gamma)} 
\left({Q^2_s\Delta^2_\perp\over 4}\right)^{\gamma}
\nonumber\\
&&
~~\times\!\left(-{2 {P^-}^2\over \Delta^2_\perp M^2_N } +i\epsilon\right)^{ \aleph(\gamma)\over 2}
(\Delta^2_\perp P^-)^{-\omega+\aleph(\gamma)}\,,
\label{Mellin}
\end{eqnarray}
with $\omega=j-1$, and where we required that the longitudinal distance (in $x^+$ direction) $L \ge \Delta^2_\perp M_N$
that is equivalent to say that our hypothetical probe has a power resolution smaller than the hadron size
\textit{i.e.} that we are in the perturbative regime.

Closing the contour to the right of $\gamma=\half$, we can now take the residue 
at the point $\gamma^*$ such that $\aleph(\gamma^*)-\omega = 0$ we have
\begin{eqnarray}
&&\hspace{-0.7cm}\int_{\Delta^2_\perp M_N}^{+\infty}\!dL \, L^{1-j}\int_{-\infty}^{+\infty}\!dy^+
\bra{P} G^{a\,i-}(L+y^+,x_\perp)
\nonumber\\
&&\hspace{-0.7cm}
~~~~~~~~~~~~~~~~~~~~~~~~~
\times[n (L +y^+) + x_\perp, ny^+ + y_\perp]^{ab}G_i^{b\,-}(y^+,y_\perp)\ket{P}
\nonumber\\
&&\hspace{-0.7cm}
= {6\,N^2_c\over \pi^2} {\Delta^{-2}_\perp \sigma_0P^-\over \aleph'(\gamma^*)}
{\gamma^*\Gamma^2(1-\gamma^*)\Gamma^3(1+\gamma^*)\over \Gamma(2+2\gamma^*)} 
\left({Q^2_s\Delta^2_\perp\over 4}\right)^{\gamma^*}
\left({2 {P^-}^2\over \Delta^2_\perp M^2_N} \right)^{ \omega\over 2}\,.
\label{Mellin-gpdfcoord0}
\end{eqnarray}
So far, the calculation was done in the ``BFKL'' limit in which ${\bar{\alpha}_s\over j-1}\simeq 1$.
To get the leading and next-to-leading residues in power of $Q^2_s\Delta^2_\perp$ we need to approach the ``DGLAP'' limit
by assuming the $\bar{\alpha}_s\ll\omega\ll1$. In this limit, then, we can use 
$\omega - \aleph(\gamma) \to \omega - {\bar{\alpha}_s \over 1-\gamma}$ 
and the leading residue is at $\gamma=1-{\bar{\alpha}_s\over \omega}$. Thus, from eq. (\ref{Mellin-gpdfcoord0}) we obtain 
\begin{eqnarray}
&&\hspace{-0.8cm}\int_{\Delta^2_\perp M_N}^{+\infty}\!dL \, L^{1-j}\int_{-\infty}^{+\infty}\!dy^+
{1\over 2P^-}\bra{P} G^{a\,i-}(L+y^+,x_\perp)
\nonumber\\
&&\hspace{-0.8cm}
~~~~~~~~~~~~~~~~~~~~~~~~~
\times\![n (L +y^+) + x_\perp, ny^+ + y_\perp]^{ab}G_i^{b\,-}(y^+,y_\perp)\ket{P}
\nonumber\\
&&\hspace{-0.8cm}
= {3\,N^2_cQ^2_s\sigma_0\over 4\pi^2}
\left({Q^2_s\Delta^2_\perp\over 4}\right)^{- {\bar{\alpha}_s\over \omega}}
\left(-{2 {P^-}^2\over \Delta^2_\perp M^2_N } +i\epsilon \right)^{ \omega\over 2}g_1(\omega)\,,
\label{Mellin-gpdfcoord}
\end{eqnarray}
where we have defined the function $g_1(\omega)$ as
\begin{eqnarray}
g_1(\omega)
\equiv {\bar{\alpha}_s\over \omega^2}{(1 - {\bar{\alpha}_s\over \omega})\Gamma^2({\bar{\alpha}_s\over \omega})
	\Gamma^3(2-{\bar{\alpha}_s\over \omega})\over \Gamma(4 - 2{\bar{\alpha}_s\over \omega})} \,.
\label{g1}
\end{eqnarray}
The first thing to notice is that we have recovered the gluon anomalous dimension ${\bar{\alpha}_s\over j-1}$ in the $j\to1$ limit
as anticipated above.
Moreover, the leading residue is leading in terms of $Q_s^2\Delta_\perp^2$ expansion, 
so eq. (\ref{Mellin-gpdfcoord}) is the leading twist contribution.

Similarly, we can calculate the next-to-leading residue in power of $\left({Q_s^2\Delta_\perp^2\over 4}\right)^{\!2}$. 
We start again from
eq. (\ref{Mellin-gpdfcoord0}) and using $\omega - \aleph(\gamma) \to \omega - {\bar{\alpha}_s \over 2-\gamma}$,
the next-to-leading residue is at $\gamma=2-{\bar{\alpha}_s\over \omega}$
\begin{eqnarray}
&&\int_{\Delta^2_\perp M_N}^{+\infty}\!dL \, L^{1-j}\int_{-\infty}^{+\infty}\!dy^+
{1\over 2P^-}\bra{P} G^{a\,i-}(L+y^+,x_\perp)
\nonumber\\
&&
~~~~~~~~~~~~~~~~~~~~~~~~~
\times[n (L +y^+) + x_\perp, ny^+ + y_\perp]^{ab}G_i^{b\,-}(y^+,y_\perp)\ket{P}
\nonumber\\
&&
\ni {3\,N^2_cQ^2_s\sigma_0\over 4\pi^2}{Q^2_s\Delta^2_\perp\over 4}
\left({Q^2_s\Delta^2_\perp\over 4}\right)^{-{\bar{\alpha}_s\over \omega}}
\left(-{2 {P^-}^2\over \Delta^2_\perp M^2_N } +i\epsilon \right)^{\!{\omega\over 2}}
g_2(\omega)\,,
\label{Mellin-gpdfcoord3}
\end{eqnarray}
where we have defined the function $g_2(\omega)$ as
\begin{eqnarray}
g_2(\omega)
\equiv {\bar{\alpha}_s\over \omega^2}{(2 - {\bar{\alpha}_s\over \omega})\Gamma^2({\bar{\alpha}_s\over \omega} - 1)
\Gamma^3(3 - {\bar{\alpha}_s\over \omega}) \over \Gamma(6 - 2{\bar{\alpha}_s\over \omega})} \,.
\label{g2}
\end{eqnarray}

We see that the next-to-leading residue, eq. (\ref{Mellin-gpdfcoord3}), in the limit of small distances $\Delta^2_\perp$,
is suppressed by one power of $Q^2_s\Delta^2_\perp\over 4$
respect to the leading residue contribution (\ref{Mellin-gpdfcoord}). 

We can continue in this way and calculate the full series of twist expansion which is resummed in eq.
(\ref{withmodel0}) or from its saddle approximation, eq. (\ref{Mpp}). However, one has to notice that, besides the 
residues which give the twist expansion, in eq. (\ref{Mellin}) there is another pole, the one at $\gamma=1$. 
This pole, actually, cancel out with two diagrams that are not included in the high-energy OPE and that have to be calculated separately.
To see how this cancellation happen, one can consider the correlation of two light-ray operator
at high-energy with ``quasi-pdf'' point splitting regulator (see section \ref{sec:analyticLR} for details). 
In this way, we may be sure that our procedure is justified. 

Adding together the leading residue, eq. (\ref{Mellin-gpdfcoord}), and the next-to-leading residue, eq. (\ref{Mellin-gpdfcoord3}), we obtain
\begin{eqnarray}
&&\hspace{-0.8cm}\int_{\Delta^2_\perp M_N}^{+\infty}\!dL \, L^{1-j}\int_{-\infty}^{+\infty}\!dy^+
{1\over 2P^-}\bra{P} G^{a\,i-}(L+y^+,x_\perp)
\nonumber\\
&&\hspace{-0.8cm}
~~~~~~~~~~~~~~~~~~~~~~~~~
\times\![n (L +y^+) + x_\perp, ny^+ + y_\perp]^{ab}G_i^{b\,-}(y^+,y_\perp)\ket{P}
\nonumber\\
&&\hspace{-0.8cm}
= {3\,N^2_cQ^2_s\sigma_0\over 4\pi^2}
\left({4\over Q^2_s\Delta^2_\perp}\right)^{{\bar{\alpha}_s\over \omega}}
\left(-{2 {P^-}^2\over \Delta^2_\perp M^2_N } +i\epsilon \right)^{ \omega\over 2}
\left(g_1(\omega) + g_2(\omega){Q^2_s\Delta^2_\perp\over 4}\right)\,.
\label{Mellin-gpdfcoordSum}
\end{eqnarray}

The final step is to perform the inverse Mellin transform of (\ref{Mellin-gpdfcoordSum})
\begin{eqnarray}
&&\hspace{-0.8cm} \int_{-\infty}^{+\infty}\!dy^+
{1\over 2P^-}\bra{P} G^{a\,i-}(L+y^+,x_\perp)
\nonumber\\
&&\hspace{-0.8cm}
~~~~~~~~~~~~~~~~~~~~~~~~~
\times\![n (L +y^+) + x_\perp, ny^+ + y_\perp]^{ab}G_i^{b\,-}(y^+,y_\perp)\ket{P}
\theta(L - \Delta^2_\perp M_N)
\nonumber\\
&&\hspace{-0.8cm}
= {3\,N^2_cQ^2_s\sigma_0\over 4\pi^2}{1\over 2\pi i}\int_{1-i\infty}^{1+i\infty}d\omega L^{\omega-1}
\left({4\over Q^2_s\Delta^2_\perp}\right)^{{\bar{\alpha}_s\over \omega}}
\left(-{2 {P^-}^2\over \Delta^2_\perp M^2_N } +i\epsilon \right)^{ \omega\over 2}
\nonumber\\
&&\hspace{-0.8cm}
~~\times\!\left(g_1(\omega) + g_2(\omega){Q^2_s\Delta^2_\perp\over 4}\right)\,.
\label{Mellin-gpdfcoordSuminv}
\end{eqnarray}
Using variables $\varrho$ and $z^2$, and eq. (\ref{lightcone-pp}) with $z^2\neq 0$, we can rewrite eq. (\ref{Mellin-gpdfcoordSuminv}) as
\begin{eqnarray}
\calm_{pp}(\varrho,z^2)
=\!\!\!&&{3\,N^2_c\over 4\pi^2}{Q^2_s\sigma_0\over \varrho}{1\over 2\pi i}\int_{1-i\infty}^{1+i\infty}d\omega\!
\left({4\over Q^2_s |z|^2}\right)^{{\bar{\alpha}_s\over \omega}}
\left({2\,\varrho^2\over z^2 M^2_N } +i\epsilon \right)^{ \omega\over 2}
\nonumber\\
&&\times\!\left(g_1(\omega) + g_2(\omega){Q^2_s|z|^2\over 4}
\right)\,.
\label{Mellin-gpdfcoordSumInv2}
\end{eqnarray}
Result (\ref{Mellin-gpdfcoordSumInv2}) is our final result which gives the 
behavior of the LT and NLT gluon distribution at high-energy in coordinate
space and from which we will calculate the pseudo and quasi-PDFs at LT and NLT corrections

In the limit $\bar{\alpha}_s\ll \omega\ll 1$, we have
\begin{eqnarray}
g_1(\omega)\simeq {1\over 6\bar{\alpha}_s}\,,~~~~~~
g_2(\omega) \simeq {2\over 15 \bar{\alpha}_s}\,.
\label{g1g2approxy}
\end{eqnarray}
So, using approximations (\ref{g1g2approxy}),
eq. (\ref{Mellin-gpdfcoordSumInv2}) becomes
\begin{eqnarray}
\calm_{pp}(\varrho,z^2)
=\!\!\!&& {N^2_c\over 8\pi^2}{Q^2_s\sigma_0\over  \bar{\alpha}_s \varrho}
{1\over 2\pi i}\int_{1-i\infty}^{1+i\infty}\!d\omega
\left({4\over Q^2_s|z|^2}\right)^{{\bar{\alpha}_s\over \omega}}
\left({2 \,\varrho^2\over z^2 M^2_N } +i\epsilon \right)^{\!{\omega\over 2}}
\nonumber\\
&&
\times\!
\left(1 + {Q^2_s|z|^2\over 5}\right)
+ O\!\left({Q^4_s|z|^4\over 16}\right)\,.
\label{Mellin-gpdfcoord-nextwist}
\end{eqnarray}
\begin{figure}[htb]
		\begin{center}
		\includegraphics[width = 2.8in]{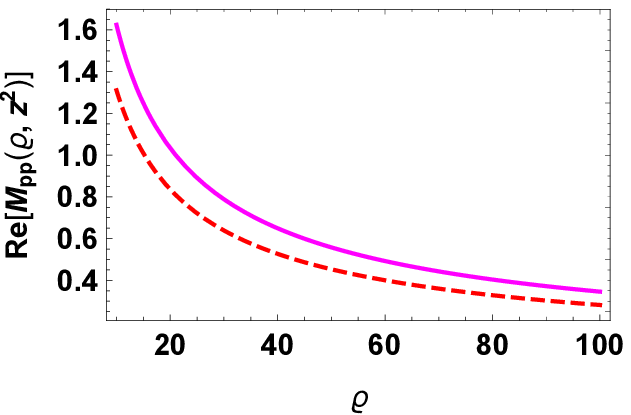}
	\includegraphics[width = 2.8in]{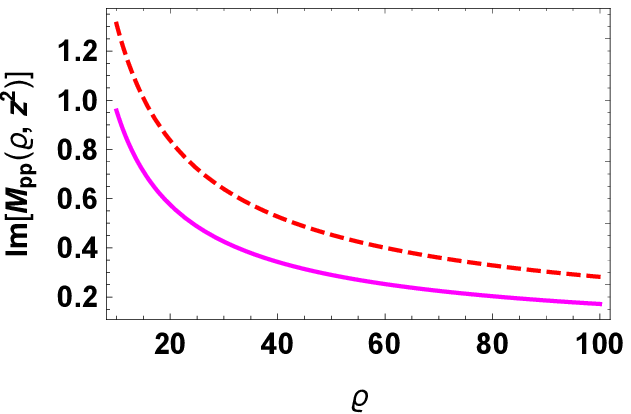}
\end{center}
	\caption{In the left and right panel we plot the real and imaginary part of eq. (\ref{Mellin-gpdfcoordSumInv2}) (red dashed curve),
		and the real and imaginary part of eq. (\ref{Bessel1gluon}) (magenta curve), respectively.}
	\label{Fig:leadnextleadappr}
\end{figure}

From eq. (\ref{Mellin-gpdfcoord-nextwist}) we can perform the inverse Mellin
analytically. We have to distinguish two cases.
In the first case, ${Q^2_s|z|^2\over 4} > 1$, the twist expansion is not justified
and all order corrections will have to be included. Indeed, if we perform the inverse Mellin transform we get
\begin{eqnarray}
\hspace{-1cm}\calm_{pp}(\varrho,z^2)
 =\!\!\!&& - {N^2_c\over 8\pi^2\bar{\alpha}_s}{Q^2_s\sigma_0\over\varrho}
\left({2\bar{\alpha}_s\ln{Q^2_s |z|^2\over 4}\over \ln\!\left({2\,\varrho^2\over z^2M_N^2}+i\epsilon\right)}\right)^{\!\!\!\half}
\!\!\! J_1(t)
\left(1 + {Q^2_s|z|^2\over 5}\right)
+ O\!\left({Q^4_s|z|^4\over 16}\right)\,,
\label{besselj1}
\end{eqnarray}
with
\begin{eqnarray}
t = \left[2\bar{\alpha}_s\ln\left({Q^2_s|z|^2\over 4}\right)
\ln\!\left({2\varrho^2\over z^2M_N^2}+i\epsilon\right)\right]^\half\,,
\end{eqnarray}
and where, we recall, $z^2<0$.
Result (\ref{besselj1}) gives an unusual behavior as a remnant of the fact that case ${Q^2_s|z|^2\over 4} > 1$
is not justified in terms of twist expansion.

For the second case $0<{Q^2_s|z|^2\over 4} <1$, 
which corresponds to the typical DIS region, the twist expansion is justified. In this case we obtain
\begin{eqnarray}
\hspace{-0.6cm}\calm_{pp}(\varrho,z^2)
= {N^2_c\over 8\pi^2\bar{\alpha}_s}{Q^2_s\sigma_0\over\varrho}
\left({4\bar{\alpha}_s \left|\ln {Q_s |z|\over 2}
\right|\over \ln\!\left({2\,\varrho^2\over z^2M_N^2}+i\epsilon\right) }\right)^{\!\!\!\half}
\!\!\! I_1(\tilde{t})
\left(1 + {Q^2_s|z|^2\over 5}\right)
+ O\!\left({Q^4_s|z|^4\over 16}\right)\,,
\label{Bessel1gluon}
\end{eqnarray}
with 
\begin{eqnarray}
\tilde{t} = \left[4\bar{\alpha}_s \left|\ln {Q_s |z|\over 2}\right|
\ln\!\left({2\,\varrho^2\over z^2M_N^2}+i\epsilon\right)\right]^\half\,.
\end{eqnarray}

\begin{figure}[htb]
	\begin{center}
		\includegraphics[width = 2.8in]{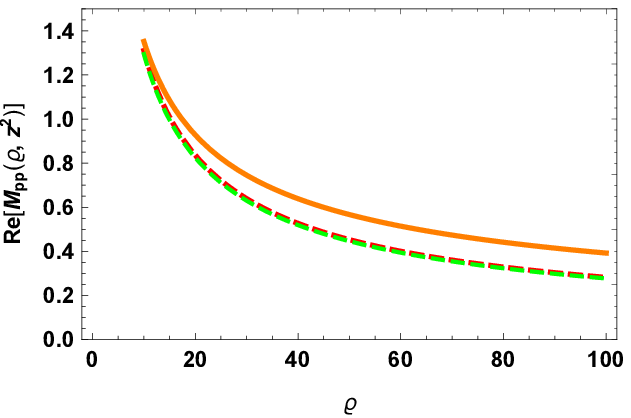}
		\includegraphics[width = 2.8in]{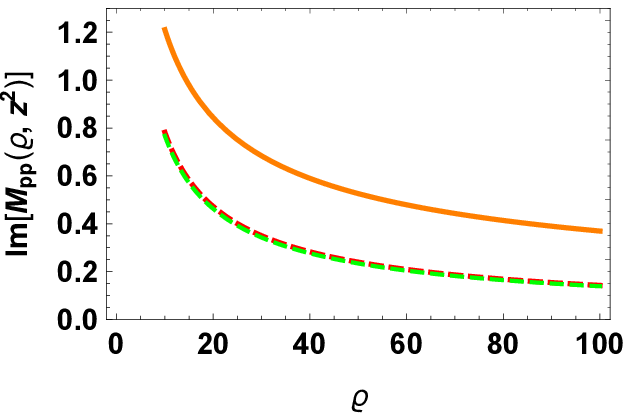}
	\end{center}
	\caption{In the left panel, the orange curve is the numerical evaluation of real part of eq. (\ref{withmodel}) (all twist resummed by BFKL), 
		the green dashed curve is the real part of the LT term in (\ref{Mellin-gpdfcoordSumInv2}) only, while the red dashed one is the 
		real part of LT+NLT result, eq. (\ref{Mellin-gpdfcoordSumInv2}).
	In the left panel we plot the imaginary parts.
		The range of $\varrho$ is from $10$ to $100$.}
	\label{Ioffeamplicompare}
\end{figure}
%{Ioffeamplicompare}
The goodness of approximation (\ref{g1g2approxy}) can be appreciated in Fig. \ref{Fig:leadnextleadappr} where we plot result
(\ref{Mellin-gpdfcoordSumInv2}) with inverse Mellin is performed numerically with $0<{Q^2_s|z|\over 4}<1$, and result (\ref{Bessel1gluon}).

\begin{figure}[htb]
	\begin{center}
		\includegraphics[width = 2.8in]{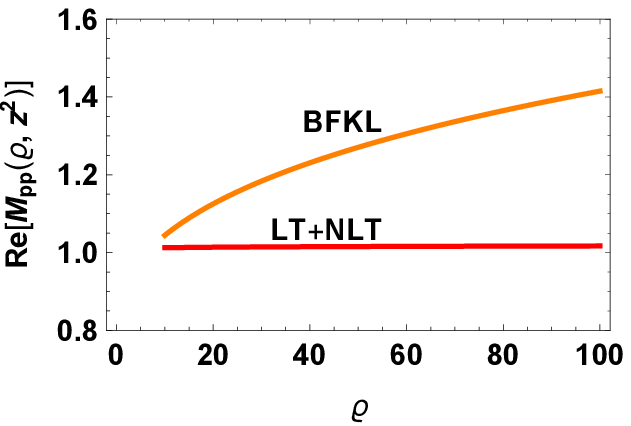}
		\includegraphics[width = 2.8in]{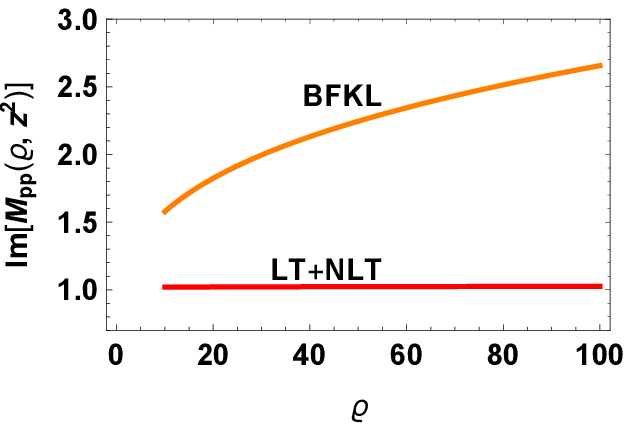}
	\end{center}
	\caption{Here we plot the real and imaginary part of the  BFKL (\ref{withmodel0}) and LT+NLT (\ref{Mellin-gpdfcoordSumInv2}) both normalized
		to the LT, respectively; $\varrho$ variates from $10$ to $100$.}
	\label{Fig:ratiosMpp}
\end{figure}

In Fig. \ref{Ioffeamplicompare} we finally compare the coordinate space result of the 
high-energy (large $\varrho$) behavior of the
gluon distribution with BFKL resummation, eq. (\ref{withmodel}), the LT term and LT plus NLT of eq. (\ref{Mellin-gpdfcoordSumInv2}).
We notice that the curves plotted in figure \ref{Ioffeamplicompare} are very slowly varying functions 
for large values of $\varrho$ (see figure \ref{Fig:ratiosMpp}).

Moreover, as mentioned in the previous section, the region of applicability of our formalism is where
the BFKL logarithm $\bar{\alpha}_s\ln\left({\sqrt{2}\varrho\over |z|M_N}\right)$ is of order 1 which is for $\varrho\ge 10$.

\section{Gluon pseudo-PDF}
\label{sec:pseudopdf}

The pseudo-PDF~\cite{Radyushkin:2017cyf} are defined as the Fourier transform of the matrix element of the gluon bi-local operator
with respect to the momentum $P^\mu$ keeping its orientation fixed.
In our notation this translates into the Fourier transform of eq. (\ref{withmodel}) and eq. 
(\ref{Mellin-gpdfcoordSumInv2}) with respect to $\varrho$.

\subsection{Gluon pseudo-PDF with BFKL resummation}

We defined the pseudo-PDF in eq. (\ref{Gp}), so we have to perform the Fourier transform of eq. (\ref{withmodel}) with respect to $\varrho$
\begin{eqnarray}
\hspace{-1.5cm}
G_{\rm p}(x_B,z^2) 
=\!\!\!&& {3N_c^2\over 4\pi^3}{Q_s\sigma_0\over |z|}\int \!\!{d\varrho\over 2\pi\varrho}e^{- i\varrho x_B}
\!\!\int\!d\nu\left({2\varrho^2\over z^2M_N^2}+i\epsilon\right)^{\!\!\aleph(\gamma)\over 2}
\nonumber\\
&&\times\!{\gamma\,\Gamma^2(1-\gamma)\Gamma^3(1+\gamma)\over \Gamma(2+2\gamma)} 
\left({|z|Q^2_s\over 2}\right)^{\!\!2i\nu}\,.
\end{eqnarray}
It is convenient to performing first the Fourier transform and the integration over the $\nu$ parameter at the end. So, we have
\begin{eqnarray}
\hspace{-1.5cm}
G_{\rm p}(x_B,z^2)
=\!\!\!&& - i{3N^2_c\over 4\pi^4}{Q_s\sigma_0\over |z|}\int\!d\nu
\left({2\over x_B^2z^2 M^2_N}+i\epsilon\right)^{\!\!\aleph(\gamma)\over 2}
\left({Q_s|z|\over 2}\right)^{2i\nu}\nonumber\\
&&\times{\gamma\,\Gamma^2(1-\gamma)\Gamma^3(1+\gamma) \Gamma(\aleph(\gamma))\over \Gamma(2+2\gamma)} 
\sin\left({\pi\over 2}\aleph(\gamma)\right){\rm sign}(x_B)\,.
\label{FourierPseudoSadl1}
\end{eqnarray}
Evaluating the last integration in the saddle-point approximation we have
\begin{eqnarray}
\hspace{-0.8cm}G_{\rm p}(x_B,z^2)
&& \simeq	- i{3N^2_c\over 128 \pi}{Q_s\sigma_0\over |z|}
{\Gamma(\bar{\alpha}_s4\ln 2)\sin({\pi\over 2}\bar{\alpha}_s4\ln 2)
	\over \sqrt{7\zeta(3)\bar{\alpha}_s\ln \left({2\over x_B^2z^2 M^2_N}+i\epsilon\right)}}	
{\rm sign}(x_B)
\nonumber\\
&&
~\times\!\exp\!\left\{{-\ln^2 {Q_s|z|\over 2}\over 7\zeta(3)\bar{\alpha}_s
\ln\left({2\over x_B^2z^2 M^2_N}+i\epsilon\right)}\right\}
\left({2\over x_B^2z^2 M^2_N}+i\epsilon\right)^{\bar{\alpha}_s2\ln 2}\,.
\label{FourierPseudoSadl2}
\end{eqnarray}
Note that, in eq. (\ref{FourierPseudoSadl2}) we can further approximate 
$\Gamma(\bar{\alpha}_s4\ln 2)\sin({\pi\over 2}\bar{\alpha}_s4\ln 2)\simeq {\pi\over 2}$ for small values of $\bar{\alpha}_s$.

\begin{figure}[htb]
	\begin{center}
		\includegraphics[width = 3.0in]{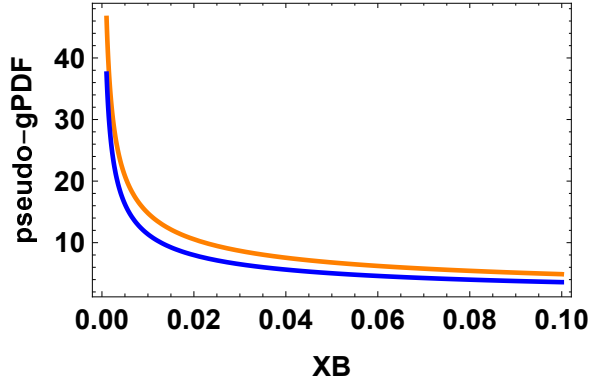}
	\end{center}
	\caption{The orange curve is the numerical integration of the real part of (\ref{FourierPseudoSadl1}), while
		the blue one is the saddle-point approximation (\ref{FourierPseudoSadl2})(real part).}
	\label{Fig:pseudo-g-sadVSnum}
\end{figure}

As can be observed in Fig. \ref{Fig:pseudo-g-sadVSnum}, the numerical evaluation of result (\ref{FourierPseudoSadl1})
is very well approximated by the saddle-point approximation result (\ref{FourierPseudoSadl2}). 

\subsection{Gluon pseudo-PDF at LT and NLT}

Let us perform the Fourier transform to obtain the pseudo-PDF for the leading and next-to-leading twist corrections.
Our starting point is the Fourier transform of eq. (\ref{Mellin-gpdfcoordSumInv2}) which, using the pseudo-PDF definition eq. (\ref{Gp}), is
\begin{eqnarray}
\hspace{-1.0cm}
G_{\rm p}(x_B,z^2)
=\!\!\!&&
{ 3N^2_c\,Q^2_s\sigma_0\over 16\pi^2}{1\over 2\pi i} \int_{1-i\infty}^{1+i\infty} d\omega 
\int_0^{+\infty}\!\!{d\varrho\over 2\pi\varrho}\, e^{-i\varrho x_B}
\nonumber\\
&&\times\!\left({2\varrho^2\over z^2M_N^2}+i\epsilon\right)^{\!\!{\omega\over 2}}
\left({4\over Q^2_s|z|^2}\right)^{\!\!{\bar{\alpha}_s\over \omega}}
\Big(4g_1(\omega) + g_2(\omega)
Q_s^2|z|^2\Big) +O\!\left({Q_s^4|z|^4\over 16}\right)\,.
\label{pseudoFT}
\end{eqnarray}
From eq. (\ref{pseudoFT}) we will first perform the Fourier transform and 
lastly the inverse Mellin transform. Thus, we have
\begin{figure}[h]

		\begin{center}
		\includegraphics[width=2.8in]{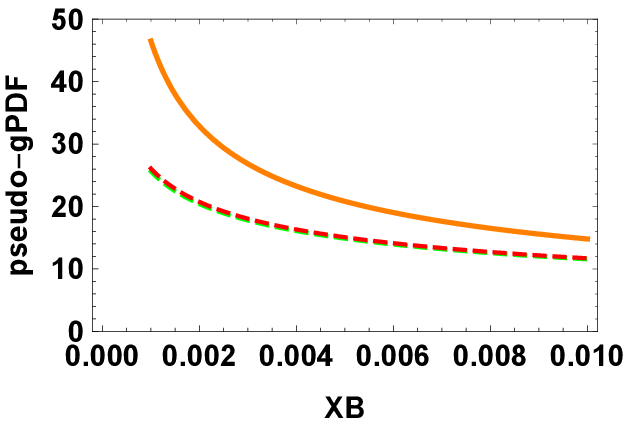}
	\includegraphics[width= 2.8in]{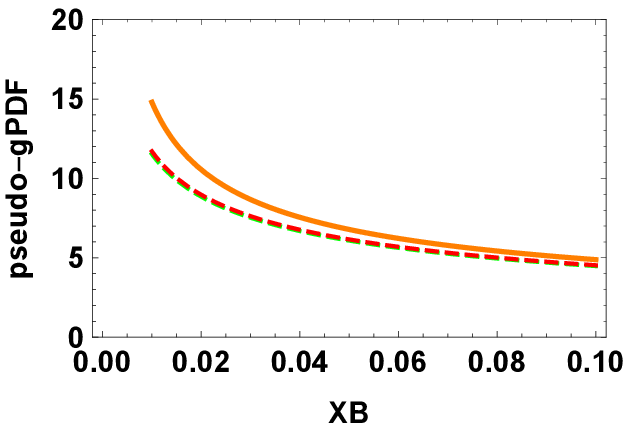}
\end{center}
	\caption{In the left panel we plot pseudo-PDF with BFKL resummation, eq. (\ref{FourierPseudoSadl1}) (orange curve), 
		and the LT (green dashed curve) and LT+NLT (red dashed curve) of pseudo-PDF result (\ref{fourier-inversmellin-g});
		The value of $x_B$ is between 0.001 to 0.02.
	In the right panel we plot the same curves in a different range of $x_B$ between 0.01 to 0.1.}
	\label{Fig:Gpseudo-comparing}
\end{figure}

\begin{eqnarray}
\hspace{-1.cm}
G_{\rm p}(x_B,z^2)
=\!\!\!&&
{3N^2_c\,Q^2_s\sigma_0\over 32\pi^3}
{1\over 2\pi i} \int_{1-i\infty}^{1+i\infty} d\omega\,
\left({2\over x_B^2|z|^2M_N^2}\right)^{\omega\over 2}
\left({4\over Q^2_s|z|^2}\right)^{{\bar{\alpha}_s\over \omega}} 	
\Gamma(\omega)
\nonumber\\
\hspace{-1.cm}
&&\times\left(4g_1(\omega) + g_2(\omega)Q_s^2|z|^2\right) +O\!\left({Q_s^4|z|^4\over 16}\right)\,.
\label{fourier-inversmellin-g}
\end{eqnarray}

The LT and NLT result, eq. (\ref{fourier-inversmellin-g}), is obtained in the limit $\bar{\alpha}_s\ll\omega\ll 1$,
so, we could further approximate  eq. (\ref{fourier-inversmellin-g}) using eq. (\ref{g1g2approxy}) and also 
approximating $\Gamma(\omega)\stackrel{\omega\to 0}{\simeq}{1\over \omega}$ and obtain
\begin{eqnarray}
\hspace{-1.cm}
G_{\rm p}(x_B,z^2)
=\!\!\!&&
{N^2_c\,Q^2_s\sigma_0\over 16\bar{\alpha}_s\pi^3}
{1\over 2\pi i} \int_{1-i\infty}^{1+i\infty} {d\omega\over \omega}\,
\left({2\over x_B^2|z|^2M_N^2}\right)^{\omega\over 2}
\left({4\over Q^2_s|z|^2}\right)^{{\bar{\alpha}_s\over \omega}} 	
\nonumber\\
\hspace{-1.cm}
&&\times\left(1 +{Q_s^2|z|^2\over 5}\right) +O\!\left({Q_s^4|z|^4\over 16}\right)\,.
\label{fourier-inversmellin-g-approxy}
\end{eqnarray}
In appendix \ref{sec:NLTanalytic-pseudoPDF} we will employ such approximations, perform the
inverse Mellin transform obtaining an analytic expression, and compare the result with the numerical evaluation
of eq. (\ref{fourier-inversmellin-g}), thus showing the goodness of the approximated analytic result. 
However, for our analysis, we will consider the numerical evaluation of eq. (\ref{fourier-inversmellin-g}).

In Fig. \ref{Fig:Gpseudo-comparing}, we plot the gluon pseudo-PDF with the BFKL resummation eq. (\ref{FourierPseudoSadl2}) (orange curve),
and the LT term and the LT plus NLT of eq. (\ref{fourier-inversmellin-g}) (green dashed and red dashed curve respectively).
We observe that the BFKL resummation result agrees with the LT and LT+NLT result in the region of moderate $x_B$.
When we move into the low-$x_B$ region, we notice a strong disagreement which confirms the necessity of 
a $\ln{1\over x_B}$ resummation represented by BFKL eq.

\section{Gluon quasi-PDF}
\label{sec:quasipdf}

%The quasi pdf \cite{Xiong:2013bka} differs from the pseudo pdf \cite{Radyushkin:2017cyf} in the Fourier transform. 
To obtain the quasi-PDF, we need to perform again a Fourier transform of eq. (\ref{withmodel}) which, this time, 
takes a slightly different form then the 
pseudo-PDF case considered in the previous section. 
Let us introduce  the real parameter $\varsigma$
such that $-z^2=\varsigma^2>0$, and the four-vector
$\xi^\mu \equiv {z^\mu\over |z|}= {z^\mu\over |\varsigma|}$ with $|z| = \sqrt{-z^2}$. 
The quasi-PDF are defined as the Fourier transform of the coordinate space gluon distributions (\ref{withmodel}) 
and (\ref{Mellin-gpdfcoordSumInv2}) keeping, this time, the orientation of the vector $z^\mu$ fixed.

\subsection{Gluon quasi-PDF with BFKL resummation}

Let us start with eq. (\ref{withmodel}).
As already mentioned before, in the high-energy limit, where $x^+$-component $\to \infty$ and
$x^-$-component $\to 0$, we can not distinguish 
between the zeroth and the third component. We can then rewrite $LP^- = z\cdot P = \varsigma P_\xi$
with $P_\xi \equiv P\cdot \xi=P^-$ because the $\xi^\mu$ vector, in the limit we are considering, selects the
minus component of the $P^\mu$ vector.
Moreover, in coordinate space, in the high-energy limit, every fields
depends only on $x^+$ and $x_\perp$, so, restoring the $x^-$ components amounts in substituting 
$(x-y)^2_\perp=\Delta^2_\perp \to -z^2 = \varsigma^2$. 
In the quasi-PDF notation eq. (\ref{withmodel0}) becomes
\begin{eqnarray}
&&\hspace{-1.5cm}
{\xi_\mu\xi_\nu\over 2P_\xi^2}\bra{P} G^{a\,\alpha\mu}(\varsigma)[\varsigma, 0]^{ab}G_\alpha^{b\,\nu}(0)\ket{P}
\nonumber\\
&&\hspace{-1.5cm}
= \,{3N^2_c\over 4\pi^3}{Q_s\sigma_0\over \varsigma^2 P_\xi}\int\!d\nu\, 
\left(-{2 P^2_\xi\over M^2_N}+i\epsilon\right)^{\aleph(\gamma)\over 2}
{\gamma\,\Gamma^2(1-\gamma)\Gamma^3(1+\gamma)\over \Gamma(2+2\gamma)} 
\left({Q^2_s\varsigma^2\over 4}\right)^{i\nu}\,.
\end{eqnarray}

We can then use the definition of the gluon quasi-PDF eq. (\ref{Gq}) 
\begin{eqnarray}
\hspace{-1.5cm}
G_{\rm q}(x_B,P_\xi) =\!\!\!&& P_\xi\int {d\varsigma\over 2\pi}\,e^{-i\varsigma P_\xi x_B}
{\xi_\mu\xi_\nu\over 2P_\xi^2}\bra{P} G^{a\,\alpha\mu}(\varsigma)[\varsigma, 0]^{ab}G_\alpha^{b\,\nu}(0)\ket{P}
\nonumber\\
=\!\!\!&& {3N^2_c\over 8\pi^4}Q_s\sigma_0
\int\!d\nu\, 
\left(-{2P^2_\xi\over M^2_N}+i\epsilon\right)^{\aleph(\gamma)\over 2}
{\gamma\,\Gamma^2(1-\gamma)\Gamma^3(1+\gamma)\over \Gamma(2+2\gamma)} 
\nonumber\\
\!\!\! &&\times\!\int\!{d\varsigma\over \varsigma^2}\left({Q^2_s\varsigma^2\over 4}\right)^{i\nu}\!e^{-i\varsigma P_\xi x_B}\,.
\end{eqnarray}
Performing the integration over $\varsigma$ we obtain (recall that we are using $\gamma = \half + i\nu$)
\begin{eqnarray}
\hspace{-1.5cm}
G_{\rm q}(x_B,P_\xi)
=\!\!\!&& i\,{3N^2_c\over 4\pi^4}Q_s \sigma_0 P_\xi |x_B|
\int\!d\nu
\left(\! - {2P^2_\xi\over M^2_N} + i\epsilon\right)^{\!\! \aleph(\gamma)\over 2}
\!\!\left(Q^2_s\over 4 P_\xi^2 x_B^2\right)^{\!\! i\nu}
\nonumber\\
&&\times\!{\gamma\,\Gamma^2(1-\gamma)\Gamma^3(1+\gamma)\Gamma(2\gamma-2)\over \Gamma(2+2\gamma)}
\sinh(\pi\nu)\,.
\label{qPDFbfkl}
\end{eqnarray}

\begin{figure}[htb]
	\begin{center}
		\includegraphics[width=3.8in]{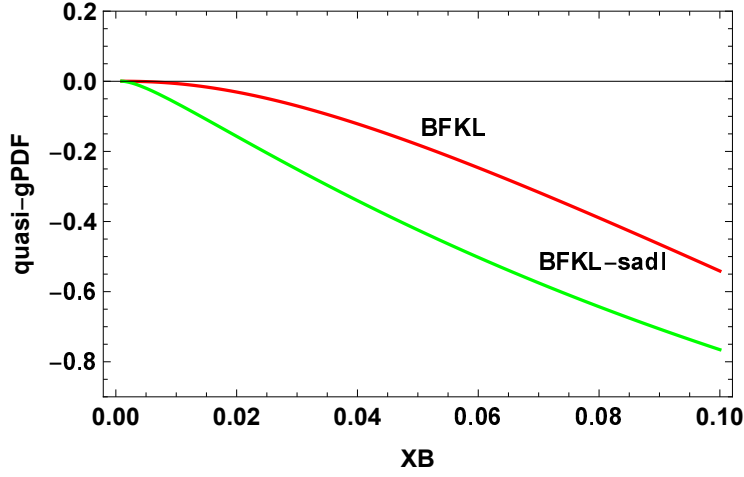}
		\caption{Here  we compare eq. (\ref{qPDFbfkl}) (real part) with its saddle point approximation, (\ref{qPDFbfklsadl}) (real part).
			The curves are plotted in the range $x_B\in [0.001,0.1]$ with $P_\xi=4$\,GeV.}
		\label{Fig:quasi-gPDFsaddle}
	\end{center}
\end{figure}

Let us calculate eq. (\ref{qPDFbfkl}) in the saddle point approximation.
To this end we note that 
${\gamma\,\Gamma^2(1-\gamma)\Gamma^3(1+\gamma)\Gamma(2\gamma-2)\over \Gamma(2+2\gamma)}
\sinh(\pi\nu)$ is a slowly varying function, so in the saddle point approximation we have
\begin{eqnarray}
	\hspace{-1.5cm}
	G_{\rm q}(x_B,P_\xi)
	\simeq
	\!\!\!&&-{3N^2_c\over 256}Q_s \sigma_0 P_\xi |x_B|
	\left(-{2P^2_\xi\over M^2_N}+i\epsilon\right)^{\bar{\alpha}_s2\ln 2}
	\!\!{e^{- {\ln^2{Q_s\over 2 P_\xi |x_B|}\over 7\bar{\alpha}_s\zeta(3)\ln\left(-{2P^2_\xi\over M^2_N}+i\epsilon\right)}}
		\over \sqrt{7\zeta(3)\bar{\alpha}_s\ln\left(-{2P^2_\xi\over M^2_N}+i\epsilon\right)}}\,.
	\label{qPDFbfklsadl}
\end{eqnarray}
In Fig. \ref{Fig:quasi-gPDFsaddle} we compare eqs. (\ref{qPDFbfklsadl}) with its saddle point approximation result (\ref{qPDFbfklsadl})
calculated with a large values of $P_\xi$. 

It is interesting to notice that, in the gluon quasi-PDF case,
the usual exponentiation of the pomeron intercept (the LO BFKL eigenvalues $\aleph(\gamma)$), which indicates the resummation of 
large logarithms of ${1\over x_B}$, is absent. 
The exponentiation of the pomeron intercept that we instead have in eq. (\ref{qPDFbfkl}) (and also in (\ref{qPDFbfklsadl}))
suggests that the logarithms resummed by BFKL equation are $\alpha_s\ln\left({2P_\xi^2\over M^2_N}\right)$
rather than the usual $\alpha_s\ln{1\over x_B}$ as in the pseudo-PDF case 
(see equations (\ref{FourierPseudoSadl1}) and (\ref{FourierPseudoSadl2})).

\subsection{Gluon quasi-PDF at LT and NLT}

Let us now consider the leading and next-to-leading twist gluon quasi-PDF. 
We have to perform the quasi-PDF Fourier transform of eq. (\ref{Mellin-gpdfcoordSuminv}) and make the inverse Mellin transform at the end. 
So, using the quasi-PDF variables $\varsigma$ and $P_\xi$, we have
\begin{eqnarray}
&&G_{\rm q}(x_B,P_\xi) 
\nonumber\\
=\!\!\!&& P_\xi\int {d\varsigma\over 2\pi}\,e^{-i\varsigma P_\xi x_B}
{\xi_\mu\xi_\nu\over 2P_\xi^2}\bra{P} G^{a\,\alpha\mu}(\varsigma)[\varsigma, 0]^{ab}G_\alpha^{b\,\nu}(0)\ket{P}
\nonumber\\
=\!\!\!&& P_\xi\int_0^{+\infty} {d\varsigma\over 2\pi}\,e^{-i\varsigma P_\xi x_B}
{3N^2_c\,Q^2_s\,\sigma_0\over 16\pi^2} 
{1\over 2\pi i} \int_{1-i\infty}^{1+i\infty} \, {d\omega \over \varsigma P_\xi}
\left(-{2P_\xi^2\over M_N^2} + i\epsilon\right)^{\omega\over 2}
\left({4\over Q^2_s\varsigma^2}\right)^{{\bar{\alpha}_s\over \omega}}
\nonumber\\
&&\times\!\Big(4g_1(\omega) + g_2(\omega)Q^2_s\varsigma^2\Big)\,.
\end{eqnarray}
Performing the integration over $\varsigma$ we get
\begin{eqnarray}
\hspace{-1.5cm}
G_{\rm q}(x_B,P_\xi)
=\!\!\!&& {3N^2_c\,Q^2_s\,\sigma_0\over 32\pi^3} 
{1\over 2\pi i} \int_{1-i\infty}^{1+i\infty}\!\! d\omega
\left(-{2P_\xi^2\over M_N^2}+i\epsilon\right)^{\!\!{\omega\over 2}}
\left(-{4P_\xi^2x_B^2\over Q^2_s}+i\epsilon\right)^{\!\!{\bar{\alpha}_s\over \omega}}
\nonumber\\
&& \times\Gamma\!\left(-{2\bar{\alpha}_s\over \omega}\right)\left(4\,g_1(\omega) 
+ {2\bar{\alpha}_s\over \omega }{Q^2_s\over P_\xi^2 x_B^2}\left(1 - {2\bar{\alpha}_s\over \omega}\right)g_2(\omega)
\right)
\label{quasiPDFtwist}\,.
\end{eqnarray}
Equation (\ref{quasiPDFtwist}) is the gluon quasi-PDF up to next-to-leading twist contribution.
What one should notice in result (\ref{quasiPDFtwist}) 
is the strong enhancement of the NLT term with respect to the LT term due to the ${1\over P_\xi^2 x_B^2}$ factor.
This is also consistent with the result obtained in Ref.~\cite{Braun:2018brg}.

\begin{figure}[htb]
	\begin{center}
		\includegraphics[width=3.8in]{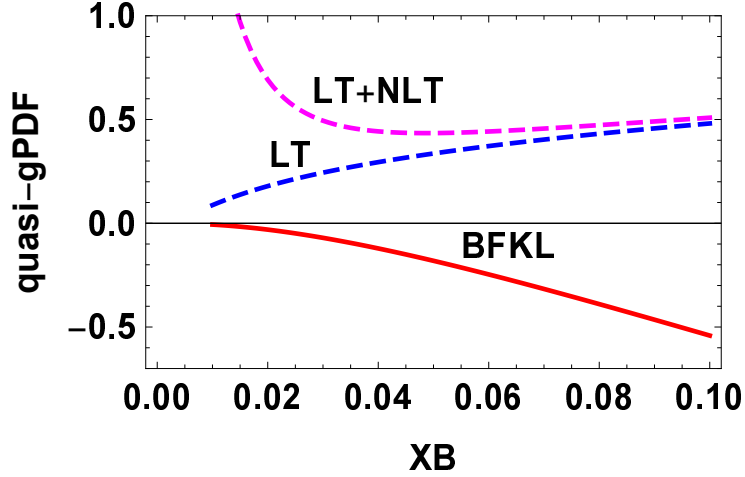}
	\end{center}
	\caption{Here we plot the 
		quasi-PDF with BFKL rsummation eq. (\ref{qPDFbfkl}) (real part), the LT and the LT+NLT of eq. (\ref{quasiPDFtwist}) (real part). 
		The curves are plotted in the range $x_B\in[0.01, 0.1]$
		with $P_\xi = 4$\,GeV.  }
	\label{Fig:quasiPDFcomparing}
\end{figure}

Since we are in the approximation $\bar{\alpha}_s\ll\omega\ll 1$, employing equations $(\ref{g1g2approxy})$, and 
$\Gamma\left(-{2\bar{\alpha}_s\over \omega}\right)\simeq - {\omega\over 2\bar{\alpha}_s}$,
we can also simplify result (\ref{quasiPDFtwist}) as
\begin{eqnarray}
	\hspace{-1.5cm}
	G_{\rm q}(x_B,P_\xi)
	=\!\!\!&& - {N^2_c\,Q^2_s\,\sigma_0\over 16\bar{\alpha}_s^2\pi^3} 
	{1\over 2\pi i} \int_{1-i\infty}^{1+i\infty}\!\! d\omega
	\left(-{2P_\xi^2\over M_N^2}+i\epsilon\right)^{\!\!{\omega\over 2}}
	\left(-{4P_\xi^2x_B^2\over Q^2_s}+i\epsilon\right)^{\!\!{\bar{\alpha}_s\over \omega}}
	\nonumber\\
	&& \times\left(\omega + {2\bar{\alpha}_sQ_s^2\over 5 P_\xi^2 x_B^2}
	+ O\left({\bar{\alpha}_s\over\omega}\right)
	\right)
	\label{quasiPDFtwistapproxy}\,.
\end{eqnarray}
In appendix \ref{sec:analytic-qPDFtwist} we provide the analytic expression of eq. (\ref{quasiPDFtwistapproxy}) in two different cases:
$\ln{4P_\xi^2x_B^2\over Q^2_s}>0$ and $\ln{4P_\xi^2x_B^2\over Q^2_s}<0$ and compare them with
result (\ref{quasiPDFtwist}) (see figures. \ref{Fig:quasiPDFtwistapproxy} and \ref{Fig:quasiPDFtwistapproxy2}).
However, although our conclusions are not affected by using either of the results (\ref{quasiPDFtwistapproxy}), or (\ref{quasiPDFtwist}),
in what follow we will plot the numerical evaluation of eq. (\ref{quasiPDFtwist}). 

In Fig. \ref{Fig:quasiPDFcomparing}, using $P_\xi=4$\,GeV, we plot the quasi-PDF with BFKL resummation (red curve) given
in eq. (\ref{qPDFbfkl}), and the LT (blue dashed curve), and LT+NLT (magenta dashed curve) contributions given in eq. (\ref{quasiPDFtwist}).
What is striking about the plot in Fig. \ref{Fig:quasiPDFcomparing} is that the behavior of the three curves is different than the usual low-$x_B$
behavior of gluon distributions which we, instead, observed for the pseudo-PDF distribution in Fig. \ref{Fig:Gpseudo-comparing}.
We considered again as initial condition of the evolution the GBW model evaluated at $x_B=0.1$. 
One may check that, changing the starting point of the evolution, for example evaluating the GBW model at $x_B=0.01$,
would not change such unusual behavior of the quasi-PDF.

\section{Conclusions}
\label{sec:conclusions}

Our findings are illustrated in figures \ref{Ioffeamplicompare}, \ref{Fig:Gpseudo-comparing}, and \ref{Fig:quasiPDFcomparing}
where we plotted the Ioffe-time distributions, the pseudo-PDF and quasi-PDF respectively.

The main result is that the pseudo-PDF and the quasi-PDF have a very different behavior at low-$x_B$. 
The physical origin of the difference between the two distributions lay in
the two different Fourier transforms under which they are defined. More
precisely, in the pseudo-PDF case, the scale is the resolution that is, the square of the 
length of the gauge link separating the bi-local operator. On
the other hand, in the quasi-PDF case, the scale is the energy that is,
the momentum of the hadronic target (the nucleon) projected along the
direction of the gauge link.
Indeed, if on one hand, the pseudo-PDF has the typical behavior  of the
gluon distribution at low-$x_B$ (see figure \ref{Fig:Gpseudo-comparing}), on the other hand,
the quasi-PDF has a rather unusual low-$x_B$ behavior (see figure \ref{Fig:quasiPDFcomparing}).
The reason is that the usual exponentiation of the BFKL pomeron intercept, which resums logarithms of $x_B$, is absent in the 
quasi-PDF result (\ref{qPDFbfkl}).
Moreover, the power corrections in the quasi-PDF result (\ref{quasiPDFtwist}) do not come in as inverse powers of $P$ but as
inverse powers of $x_B P$, so for low values of $x_B$ and fixed values of $P$ these corrections 
are enhanced rather than suppressed at this regime.

Another result we obtained is the large-distance behavior of the gluon Ioffe-time distribution 
(see figure \ref{Ioffeamplicompare}) where we noticed that the plotted curves (for the real and imaginary part) are 
very slowly varying functions for large values of $\varrho$ (see figure \ref{Fig:ratiosMpp}).
Indeed, since in lattice calculations the values of $\varrho$ is not very large, to perform the Fourier transform and
obtain the $x_B$ dependence,
one has to extrapolate the large-distance behavior of the Ioffe-time distribution. 

In this work, running coupling corrections and next-to-leading order BFKL have not been used and could be
included. 
Moreover, the technique we developed in this work can be extended to study the low-$x_B$ behavior of
other unpolarized or polarized pseudo and quasi parton distributions.

The author is grateful to I. Balitsky and V. Braun for valuable discussions.
He also thanks A. Manashov and A. Vladimirov for discussions.

\appendix

\section{Projection with open transverse indexes}
\label{sec:projection}

He we consider the gluon matrix element with open transverse indexes. We start from
\begin{eqnarray}
&&\hspace{-2cm}\langle G^{a\,i-}(x^+,x_\perp)[n x^+ + x_\perp, ny^+ + y_\perp]^{ab}G^{b\,j-}(y^+,y_\perp)
\rangle_{{\rm Fig.}\ref{Fig:quasipdf-gq-LO}}
\nonumber\\
&&\hspace{-2cm}
= (\partial_x^i g^{\mu -} - \partial_x^-g^{i\mu})(\partial_y^j g^{\nu -} - \partial_y^-g^{j\nu})
\langle A^a_\mu(x)A^b_\nu(y)\rangle_A\,.
\label{before-gluonprop}
\end{eqnarray}
Using the gluon propagator (\ref{gluonprop-coord}) in (\ref{before-gluonprop}) we get
\begin{eqnarray}
&&\hspace{-2cm}
\langle G^{a\,i-}(x^+,x_\perp)[n x^+ + x_\perp, ny^+ + y_\perp]^{ab}G^{b\,j-}(y^+,y_\perp)
\rangle_{{\rm Fig.}\ref{Fig:quasipdf-gq-LO}}
\nonumber\\
&&\hspace{-2cm}
= (\partial_x^-g^{i\mu} - \partial_x^i g^{\mu -})(\partial_y^j g^{\nu -} - \partial_y^-g^{j\nu})
\nonumber\\
&&\hspace{-2cm}
\times\!
\int {d^2z_2\over 4\pi^3}\, {x^+|y^+|g^\perp_{\mu\nu} - |y^+|n'_{\mu}X^\perp_{2\nu} 
+ x^+n'_{\nu}Y^\perp_{2\mu} + n'_{\mu}n_{2\nu}  (X_2,Y_2)\over 
\Big[|y^+|(x-z_2)^2_\perp + x^+(y-z_2)^2_\perp - \ie\Big]^2}\,U^{ab}_{z_2}U^{ab}_{z_1}\,.
\label{after-gluonprop}
\end{eqnarray}
After differentiation eq. (\ref{after-gluonprop}) becomes
\begin{eqnarray}
&&\hspace{-0.8cm}\langle G^{a\,i-}(x^+,x_\perp)[n x^+ + x_\perp, ny^+ + y_\perp]^{ab}G^{b\,j-}(y^+,y_\perp)
\rangle_{{\rm Fig.}\ref{Fig:quasipdf-gq-LO}}
\nonumber\\
&&\hspace{-0.8cm}
= \int{d^2z_2\over 4\pi^3}\Bigg\{ 
{4\,g^{ij}\over [x^+(y-z_2)^2_\perp-y^+(x-z_2)^2_\perp]^2}
\nonumber\\
&&\hspace{-0.8cm}
+ {4y^+g^{ij}(x-z_2)^2_\perp - 4 x^+g^{ij}(y-z_2)^2_\perp - 12 y^+(x-z_2)^i(x-z_2)^j + 12 x^+(y-z_2)^i(y-z_2)^j
	\over [x^+(y-z_2)^2_\perp-y^+(x-z_2)^2_\perp]^3}
\nonumber\\
&&\hspace{-0.8cm}
+ {1\over [x^+(y-z_2)^2_\perp-y^+(x-z_2)^2_\perp]^4}\Big(
24 x^+y^+(x-z_2,y-z_2)(x-z_2)^i(y-z_2)^j 
\nonumber\\
&&\hspace{-0.8cm}
- 12{y^+}^2(x-z_2)^2_\perp (x-z_2)^i(x-z_2)^j - 12{x^+}^2(y-z_2)^2_\perp (y-z_2)^i(y-z_2)^j 
\nonumber\\
&&\hspace{-0.8cm}
-6 g^{ij}x^+y^+(x-z_2)^2_\perp(y-z_2)^2_\perp\Big)\Bigg\}U^{ab}_{z_2}U^{ab}_{z_1}\,.
\label{after-gluonprop2}
\end{eqnarray}

We recall that the point $z_1$ can be anywhere between point $nx^++x_\perp$ and $ny^++y_\perp$.
We parametrize the straight line between these two points as 
$x_u = ux_\perp + \baru y_\perp = z_{1\perp}$, with $u = {|y^+|\over \Delta^+}$, $\baru = {x^+\over \Delta^+}$
and $\Delta^+ = x^+ - y^+$, and
at the end we will have to integrate over the parameter $u$. 

Considering forward matrix elements and including the solution of the linear evolution of the Wilson-line operator (BFKL equation)
\begin{eqnarray}
&&\hspace{-1cm}\calv^a(z_{12}) = \!\!\int{d\nu\over  2\pi^2}
\,(z_{12}^2)^{-\half+i\nu}\left(- {2x^+ y^+\over (x-y)^2a_0}+i\epsilon\right)^{{\alpha_s N_c\over 2\pi}\chi(\gamma)}
\!\!\int d^2\omega (\omega^2_\perp)^{-\half-i\nu}\calv^{a_0}(\omega_\perp)
\end{eqnarray}
with ${1\over z_{12}^2}\calu^a(z_{12}) = \calv^a(z_{12})$, $\aleph(\gamma)= \bar{\alpha}_s\chi(\gamma)$, $\gamma = \half+i\nu$, and
$\chi(\gamma) = 2\psi(1) - \psi(\gamma) - \psi(1-\gamma)$,
from (\ref{after-gluonprop2}) we arrive at
\begin{eqnarray}
&&\hspace{-0.8cm}\langle G^{a\,i-}(x^+,x_\perp)[n x^+ + x_\perp, ny^+ + y_\perp]^{ab}G^{b\,j-}(y^+,y_\perp)
\rangle_{{\rm Fig.}\ref{Fig:quasipdf-gq-LO}}
\nonumber\\
&&\hspace{-0.8cm}
= \int{d^2z_2\over 4\pi^3}\Bigg\{ 
{4\,g^{ij}\over [x^+(y-z_2)^2_\perp-y^+(x-z_2)^2_\perp]^2}
\nonumber\\
&&\hspace{-0.8cm}
~~+ {4y^+g^{ij}(x-z_2)^2_\perp - 4 x^+g^{ij}(y-z_2)^2_\perp - 12 y^+(x-z_2)^i(x-z_2)^j + 12 x^+(y-z_2)^i(y-z_2)^j
	\over [x^+(y-z_2)^2_\perp-y^+(x-z_2)^2_\perp]^3}
\nonumber\\
&&\hspace{-0.8cm}
~~+ {1\over [x^+(y-z_2)^2_\perp-y^+(x-z_2)^2_\perp]^4}\Big(
24 x^+y^+(x-z_2,y-z_2)(x-z_2)^i(y-z_2)^j 
\nonumber\\
&&\hspace{-0.8cm}
~~- 12{y^+}^2(x-z_2)^2_\perp (x-z_2)^i(x-z_2)^j - 12{x^+}^2(y-z_2)^2_\perp (y-z_2)^i(y-z_2)^j 
\nonumber\\
&&\hspace{-0.8cm}
~~-6 g^{ij}x^+y^+(x-z_2)^2_\perp(y-z_2)^2_\perp\Big)\Bigg\}
\nonumber\\
&&\hspace{-0.8cm}
~~\times\! z_{12}^2
\int{d\nu\over  2\pi^2}
\,(z_{12}^2)^{-\half+i\nu}\left(- {2x^+ y^+\over (x-y)^2a_0}+i\epsilon\right)^{{\alpha_s N_c\over 2\pi}\chi(\gamma)}
\int d^2\omega (\omega^2_\perp)^{-\half-i\nu}\calv^{a_0}(\omega_\perp)\,.
\label{after-gluonprop3}
\end{eqnarray}

The projection over the open indexes tensor structure is
\begin{eqnarray}
&&\hspace{-0.8cm}\int{d^2z_2\over 4\pi^3}\Bigg\{ 
{4\,g^{ij}\over [x^+(y-z_2)^2_\perp-y^+(x-z_2)^2_\perp]^2}
\nonumber\\
&&\hspace{-0.8cm}
+ {4y^+g^{ij}(x-z_2)^2_\perp - 4 x^+g^{ij}(y-z_2)^2_\perp - 12 y^+(x-z_2)^i(x-z_2)^j + 12 x^+(y-z_2)^i(y-z_2)^j
	\over [x^+(y-z_2)^2_\perp-y^+(x-z_2)^2_\perp]^3}
\nonumber\\
&&\hspace{-0.8cm}
+ {1\over [x^+(y-z_2)^2_\perp-y^+(x-z_2)^2_\perp]^4}\Big(
24 x^+y^+(x-z_2,y-z_2)(x-z_2)^i(y-z_2)^j 
\nonumber\\
&&\hspace{-0.8cm}
- 12{y^+}^2(x-z_2)^2_\perp (x-z_2)^i(x-z_2)^j - 12{x^+}^2(y-z_2)^2_\perp (y-z_2)^i(y-z_2)^j 
\nonumber\\
&&\hspace{-0.8cm}
-6 g^{ij}x^+y^+(x-z_2)^2_\perp(y-z_2)^2_\perp\Big)\Bigg\}(z^2_{12})^\gamma
\nonumber\\
&&\hspace{-0.8cm}
= g^{ij} {3\over 2}
{\gamma^2\Gamma(1+\gamma)\Gamma(1-\gamma)\over \pi^2 {\Delta^+}^2}{(u\baru)^{\gamma} \over [ \Delta^2_\perp]^{1-\gamma}}\,,
\label{final-openindex-pojection}
\end{eqnarray}
with $\Delta_\perp^2 = (x-y)^2_\perp = - (x-y)^i(x-y)_i$. We see that only one tensor structure survived after projection.
So, using result (\ref{final-openindex-pojection}) in eq. (\ref{after-gluonprop3}) we obtain
\begin{eqnarray}
&&\hspace{-1cm}\calv^a(z_{12}) = \int{d\nu\over  2\pi^2}
\,(z_{12}^2)^{-\half+i\nu}\left(- {2x^+ y^+\over (x-y)^2a_0}+i\epsilon\right)^{{\alpha_s N_c\over 2\pi}\chi(\gamma)}
\!\!\int d^2\omega (\omega^2_\perp)^{-\half-i\nu}\calv^{a_0}(\omega_\perp)\,,
\end{eqnarray}
with ${1\over z_{12}^2}\calu^a(z_{12}) = \calv^a(z_{12})$, $\aleph(\gamma)= \bar{\alpha}_s\chi(\gamma)$, $\gamma = \half+i\nu$, and
$\chi(\gamma) = 2\psi(1) - \psi(\gamma) - \psi(1-\gamma)$,
from (\ref{after-gluonprop2}) we arrive at
\begin{eqnarray}
&&\hspace{-0.8cm}\langle G^{a\,i-}(x^+,x_\perp)[n x^+ + x_\perp, ny^+ + y_\perp]^{ab}G^{b\,j-}(y^+,y_\perp)
\rangle_{{\rm Fig.}\ref{Fig:quasipdf-gq-LO}}
\nonumber\\
&&\hspace{-0.8cm}
= \int{d\nu\over  2\pi^2}
\left(- {2x^+ y^+\over (x-y)^2a_0}+i\epsilon\right)^{{\alpha_s N_c\over 2\pi}\chi(\gamma)}
\int d^2\omega (\omega^2_\perp)^{-\half-i\nu}\calv^{a_0}(\omega_\perp)
\nonumber\\
&&\hspace{-0.8cm}
~\times 
g^{ij} {3\over 2}
{\gamma^2\Gamma(1+\gamma)\Gamma(1-\gamma)\over \pi^2 {\Delta^+}^2}{(u\baru)^{\gamma} \over [ \Delta^2_\perp]^{1-\gamma}}
\,.
\label{after-gluonprop4}
\end{eqnarray}
Integrating over $u$ and contracting the transverse indexes $i$ and $j$, from (\ref{after-gluonprop4}) we arrive at result (\ref{projection}).

\section{From local operators to Light-ray operators}
\label{sec:analyticLR}

It is known that correlation functions of non-local operators on the light-cone are UV divergent in the high-energy (Regge) limit
and that a way to regulate these divergences is to consider the point-splitting regulator.
In this section we will first show that the ``quasi-pdf frame'' (see figure \ref{frames}) is a valid point-splitting regulator 
for correlation function of non-local operators at high-energy (Regge) limit, and that, in this limit, it
gives the same result as the one obtained in Refs. \cite{Balitsky:2015oux, Balitsky:2018irv} using the ``Wilson-frame''
(see figure \ref{frames}). In this way we can show that diagrams a) and b) of figure \ref{Fig:first3ordersDiagrams}
 which are not included in the HE-OPE formalism do not contribute. This is because these diagrams cancel out with the
 residue at $\gamma=1$ as explained in section \ref{sec:LT-NLT}.

In this section we will also use light-cone vectors $p_1$ and $p_2$ such that $2p_1\cdot p_2 = s$,
and $x_{p_1}= p_1^\mu x_\mu = \sqrt{s/2}\,x^-$, and $x_{p_2}= p_2^\mu x_\mu = \sqrt{s/2}\,x^+$.
The light cone vectors $n^\mu$ and $n'^\mu$ that we introduced in section
\ref{sec:defoperator} are related to the light-cone vectors $p_1$ and $p_2$ by $p_1^\mu=\sqrt{s\over 2}n^\mu$
and $p_2^\mu = \sqrt{s\over 2}n'^\mu$.

\subsection{Analytic continuation of local twist-two operator to non-integer spin $j$}
\label{sec:analyticlocal}

Using the Hankel representation of the Gamma function we can write
\begin{eqnarray}
&&\hspace{-1cm}{1\over \Gamma(j-1)}F_{p_1\xi}^a(x)\nabla^{j-2}_{p_1}F_{p_1}^{a~\xi}(x)\Big|_{x=0} \nonumber\\
&&=  - {1\over 2\pi i}
\int_{H_+}\!\!du\,(-u)^{1-j}\,F_{p_1\xi}^a(x)e^{-u\nabla_{p_1}}F_{p_1}^{a~\xi}(x)\Big|_{x=0}
\nonumber\\
&&= - {1\over 2\pi i}
\int_{H_+}\!\!du\,(-u)^{1-j}\,F_{p_1\xi}^a(0)[0,-up_1]^{ab}F_{p_1}^{b~\xi}(-up_1)\,,
\label{Hankelplus}
\end{eqnarray}
with $j-1\in \mathbb{C}$, and where $H_+$ is the Hankel contour which starts at $+\infty$ slightly above the real axis,
goes around the origin counter-clockwise and goes back to $+\infty$ slightly below the real axes.

Using the $H_-$ Hankel contour, which starts at $-\infty$ slightly below the real axis, goes around the origin counter-clockwise and goes back
to $-\infty$ slightly above the real axis, we can rewrite eq. (\ref{Hankelplus}) as
\begin{eqnarray}
{1\over \Gamma(j-1)}F_{p_1\xi}^a(x)\nabla^{j-2}_{p_1}F_{p_1}^{a~\xi}(x)\Big|_{x=0} 
=  {1\over 2\pi i}
\int_{H_-}\!\!du\,\,u^{1-j}\,F_{p_1\xi}^a(up_1)[up_1,0]^{ab}F_{p_1}^{b~\xi}(0)\,,
\label{twist2gamma}
\end{eqnarray}
with $(j-1)\in \mathbb{C}$. 
Now, let us consider $j \in \mathbb{C} - \{1,2,3,4,\dots\}$ in (\ref{twist2gamma}), then 
we can leave the Hankel contour and obtain
\begin{eqnarray}
&&\hspace{-0.7cm}{1\over \Gamma(j-1)}F_{p_1\xi}^a(x)\nabla^{j-2}_{p_1}F_{p_1}^{a~\xi}(x)\Big|_{x=0} 
\nonumber\\
&&={\sin[\pi(j-1)]\over \pi}\int^{\infty}_0 dv\, v^{1-j}\, F^a_{p_1\xi}(-vp_1)[-vp_1,0]^{ab}F^{b~\xi}_n(0)\,,
\label{integratingHminus}
\end{eqnarray}
with  $j \in \mathbb{C} - \{1,2,3,4,\dots\}$. 
Since we are interested in the forward matrix elements we can rewrite (\ref{integratingHminus}) as
\begin{eqnarray}
&&\hspace{-0.7cm}{1\over \Gamma(j-1)}F_{p_1\xi}^a(x)\nabla^{j-2}_{p_1}F_{p_1}^{a~\xi}(x)\Big|_{x=0} 
\nonumber\\
&&\stackrel{\rm forw.}{=}{\sin[\pi(j-1)]\over \pi}\int^{\infty}_0 dv\, v^{1-j}\, F^a_{p_1\xi}(0)[0,vp_1]^{ab}F^{b~\xi}_n(vp_1)\,.
\label{integratingHminus2}
\end{eqnarray}
Using the reciprocal formula of Gamma function we can finally write (\ref{integratingHminus2}) as
\begin{eqnarray}
\hspace{-0.6cm}F_{n\xi}^a(x)\nabla_n^{j-2}F_{p_1}^{a~\xi}(x)\Big|_{x=0} 
\stackrel{\rm forw.}{=}  {1\over \Gamma(2-j)}\int^{\infty}_0 dv\, v^{1-j}\, F^a_{p_1\xi}(0)[0,vp_1]^{ab}F^{b~\xi}_{p_1}(vp_1)\,.
\label{integratingHminus-forw-b}
\end{eqnarray}
Equation (\ref{integratingHminus-forw-b}) is the analytical continuation of the twist-two gluon operator to non-integer values of $j$
for forward matrix elements.

Similarly, if we consider the scalar twist-two operator we get
\begin{eqnarray}
\bar{\phi}_{AB}^a(x)\nabla^j_{p_1}\phi^{ABa}(x)\Big|_{x=0}
\stackrel{\rm forw.}{=}
{1\over \Gamma(-j)}\int_0^{+\infty}\!dv\,v^{-1-j}\bar{\phi}_{AB}^a(vp_1)[vp_1,0]^{ab}\phi^{ABb~\xi}(0)\,,
\label{twis2Hankelscalar}
\end{eqnarray}
and for the gluino twist-two operator we get
\begin{eqnarray}
\hspace{-1cm}i\bar{\lambda}^a_A(x)\nabla^{j-1}_{p_1}\sigma_{p_1}\lambda_A^a(x)|_{x=0} 
\stackrel{\rm forw.}{=}
{1\over \Gamma(1-j)}\int_0^{+\infty}\!\!dv\,\,v^{-j}\,{i\over 2}&&\Big[-\bar{\lambda}^a_A(0)[0,vp_1]^{ab}\sigma_{p_1}\lambda^b_A(vp_1)
\nonumber\\
&&+\bar{\lambda}^a_A(vp_1)[vp_1,0]^{ab}\sigma_{p_1}\lambda_A^b(0)\Big]\,.
\label{twis2Hankelgluino}
\end{eqnarray}

\subsection{Super-multiplet of local operators in CFT}

Let us consider the super-multiplet of local operators \cite{Belitsky:2003sh} defined as \cite{Balitsky:2018irv}
\begin{eqnarray}
&&\calo^j_\phi (x_\perp) = \int\!du\, \bar{\phi}^a_{AB}\nabla_{p_1}^j\phi^{ABa}(up_1+x_\perp)\,,
\label{Ojscalar}\\
&&\calo^j_\lambda (x_\perp) = \int\!du\, i\bar{\lambda}^a_A\nabla_{p_1}^{j-1}\lambda^a_A(up_1+x_\perp)\,,
\label{Ojgluino}\\
&&\calo^j_g (x_\perp) = \int\!du\, F^a_{p_1 i}\nabla_{p_1}^{j-2}{{F^a}_{p_1}}^i(up_1+x_\perp)\,.
\label{Ojgluon}
\end{eqnarray}
In the case of forward matrix elements the multiplicatively renormalizable operators are 

\begin{eqnarray}
&&S_1^j = \calo_g^j + {1\over 4}\calo^j_\lambda - \half \calo^j_\phi\,,
\label{S1fForwardOj}
\\
&&S^j_2 = \calo^j_g - {1\over 4(j-1)}\calo^j_\lambda + {j+1\over 6(j-1)}\calo^j_\phi\,,
\label{S2fForwardOj}
\\
&&S^j_3 = \calo^j_g - {j+2\over 2(j-1)}\calo^j_\lambda - {(j+1)(j+2)\over 2j(j-1)}\calo^j_\phi\,,
\label{S3fForwardOj}
\end{eqnarray}
with anomalous dimensions \cite{Belitsky:2003sh}
\begin{eqnarray}
\gamma_j^{S_1} = 4[\psi(j-1)+\gamma_E] + O(\alpha_s^2), 
~~~\gamma_j^{S_2} = \gamma_{j+2}, ~~~\gamma^{S_3}_j = \gamma^{S_1}_{j+4}\,.
\end{eqnarray}

In conformal field theory, the two-point correlation function is determined by symmetry up to a coefficient, the
structure constant.
If we consider two operators $\calo^j_{p_1}(x)$ and $\calo^{j'}_{p_2}(x)$ of spin-$j$ and spin-$j'$
and indexes contracted with two light-like vectors $p_1$ and $p_2$, respectively,
then the two-point correlation function can be written as

\begin{eqnarray}
	\int dv du\langle\calo^j_{p_1}(up_1+x_\perp)\calo^{j'}_{p_2}(vp_2+y_\perp)\rangle
	= \delta(j-j'){C(\Delta,j)s^{j-1}\over [(x-y)^2_\perp]^{\Delta-1}}\mu^{-2\gamma_a}\,,
	\label{2point-general}
\end{eqnarray}
where $\Delta = d+\gamma_a$ with $d$ canonical dimension of the operator, $\gamma_a$
the anomalous dimension, $s=2p_1\cdot p_2$, $\mu$ is the renormalization point,
and $C(\Delta,j)$ is the structure constant. 

\subsection{Super-multiplet of non-local twist-two operator with non-integer spin $j$}

Following the procedure explained in section \ref{sec:analyticlocal}
we can obtain the analytical continuation of the super-multiplet local operators
(\ref{S1fForwardOj}), (\ref{S2fForwardOj}), and (\ref{S3fForwardOj})
to non-integer $j$
\begin{eqnarray}
&&\calf^j_{p_1}(x_\perp) = \int_0^{\infty}\! du\, u^{1-j}\calf_{p_1}(up_1+x_\perp)\,,\\
&&\Lambda^j_{p_1}(x_\perp) = \int_0^{\infty}\! du\, u^{-j}\Lambda_{p_1}(up_1+x_\perp)\,,\\
&&\Phi^j_{p_1}(x_\perp) = \int_0^\infty\! du\, u^{-1-j}\Phi_{p_1}(up_1+x_\perp)\,,
\end{eqnarray}
with
\begin{eqnarray}
&&\calf^j_{p_1}(up_1,x_\perp) = \int \!dv 
\,{F^a}_{p_1\mu}(up_1+vp_1 + x_\perp)[up_1+vp_1,vp_1]_x^{ab}{{F^b}_{p_1}}^\mu(vp_1+x_\perp)\,,\\
&&\Lambda^j_{p_1}(up_1,x_\perp) = {i\over 2}\int \!dv\Big(-\bar{\lambda}^a_A(up_1+vp_1 + x_\perp)
[up_1+vp_1,vp_1]_x^{ab}\sigma_-\lambda^b_A(vp_1+x_\perp)
\nonumber\\
&&~~~~~~~~~~~~~~~~~~~~~ + \bar{\lambda}^a_A(vp_1 + x_\perp)
[vp_1, up_1+vp_1]_x^{ab}\sigma_{p_1}\lambda_A^b(up_1+vp_1+x_\perp)\Big)\,,\\
&&\Phi^j_{p_1}(u,x_\perp) = \int\!dv\,\phi_I^a(up_1+vp_1+x_\perp)[up_1+vp_1,vp_1]_x^{ab}\phi_I^b(vp_1+x_\perp) \,.
\end{eqnarray}
Thus, the analytic continuation of the multiplicatively renormalizable light-ray operators to non-integer $j$ are

\begin{eqnarray}
&&\cals_1 = \calf^j_{p_1} + {j-1\over 4}\Lambda^j_{p_1} - j(j-1)\half \Phi^j_{p_1}\,,
\label{S1Forwardj}
\\
&&
\cals_2 =  \calf^j_{p_1} - {1\over 4}\Lambda^j_{p_1} + {j(j+1)\over 6}\Phi^j_{p_1}\,,
\label{S2Forwardj}
\\
&&\cals_3 =  \calf^j_{p_1} - {j+2\over 2}\Lambda^j_{p_1} - {(j+1)(j+2)\over 2}\Phi^j_{p_1}\,.
\label{S3Forwardj}
\end{eqnarray}
Notice that, the different coefficients between the $S$-operators in (\ref{S1fForwardOj})-(\ref{S3fForwardOj})
and the $\cals$-operators in (\ref{S1Forwardj})-(\ref{S3Forwardj}) are due to 
eqs. (\ref{integratingHminus-forw-b}), (\ref{twis2Hankelscalar}), and (\ref{twis2Hankelgluino})

For the two-point correlation function constructed with the $\cals$-operaotrs, it holds similar general result
\begin{eqnarray}
\langle S^j(x_\perp) S^{j'}(y_\perp)\rangle = \delta(j-j'){C(\Delta,j)s^{j-1}\over [(x-y)^2_\perp]^{\Delta-1}}\mu^{-2\gamma_a}\,.
\label{2pointS}
\end{eqnarray}

We will calculate the $C(\Delta,j)$ in the BFKL limit, \textit{i.e.} in the 
$\omega = j-1\to 0$, coupling constant $g\to 0$ and ${g\over\omega}\sim 1$ and then we will also consider the
limit $g^2\ll\omega\ll1$.

From (\ref{S1Forwardj})-(\ref{S3Forwardj}) we have that
\begin{eqnarray}
&&\calf^j_{p_1} = {(1+j)(2+j)\over 6j^2}\cals_1^j - {(1-j)(2+j)(1+3j)\over 2 j^2(3+2j)}\cals_2^j + {(1-j)(2-j)\over 6j(3+2j)}\cals_3^j\,.
\end{eqnarray}
So, we deduce that, in the BFKL limit, calculating the two-point correlation function of $\calf^j$ si 
equivalent to calculate the two-point correlation function of $\cals_1^j$ for which it holds eq. (\ref{2pointS}).
\begin{figure}
\begin{center}
\includegraphics[width=63mm]{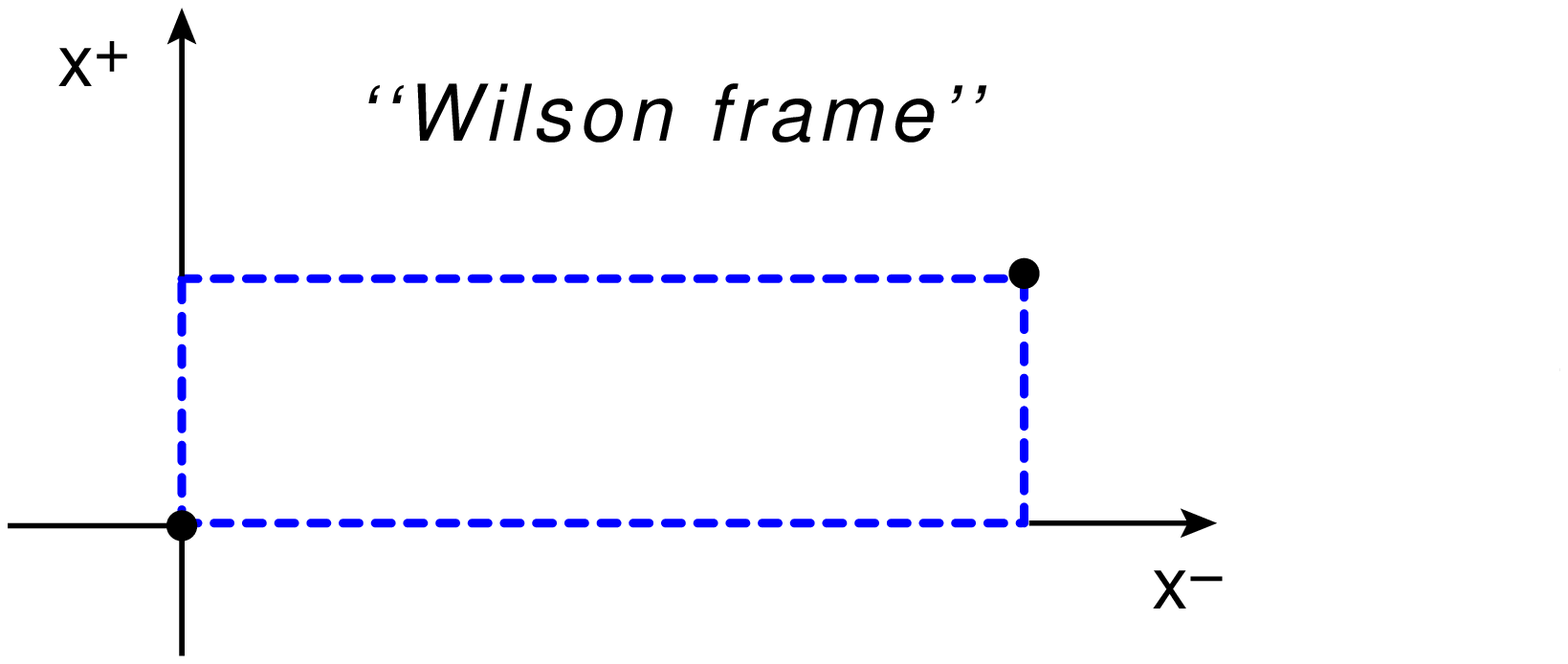}
\includegraphics[width=43mm]{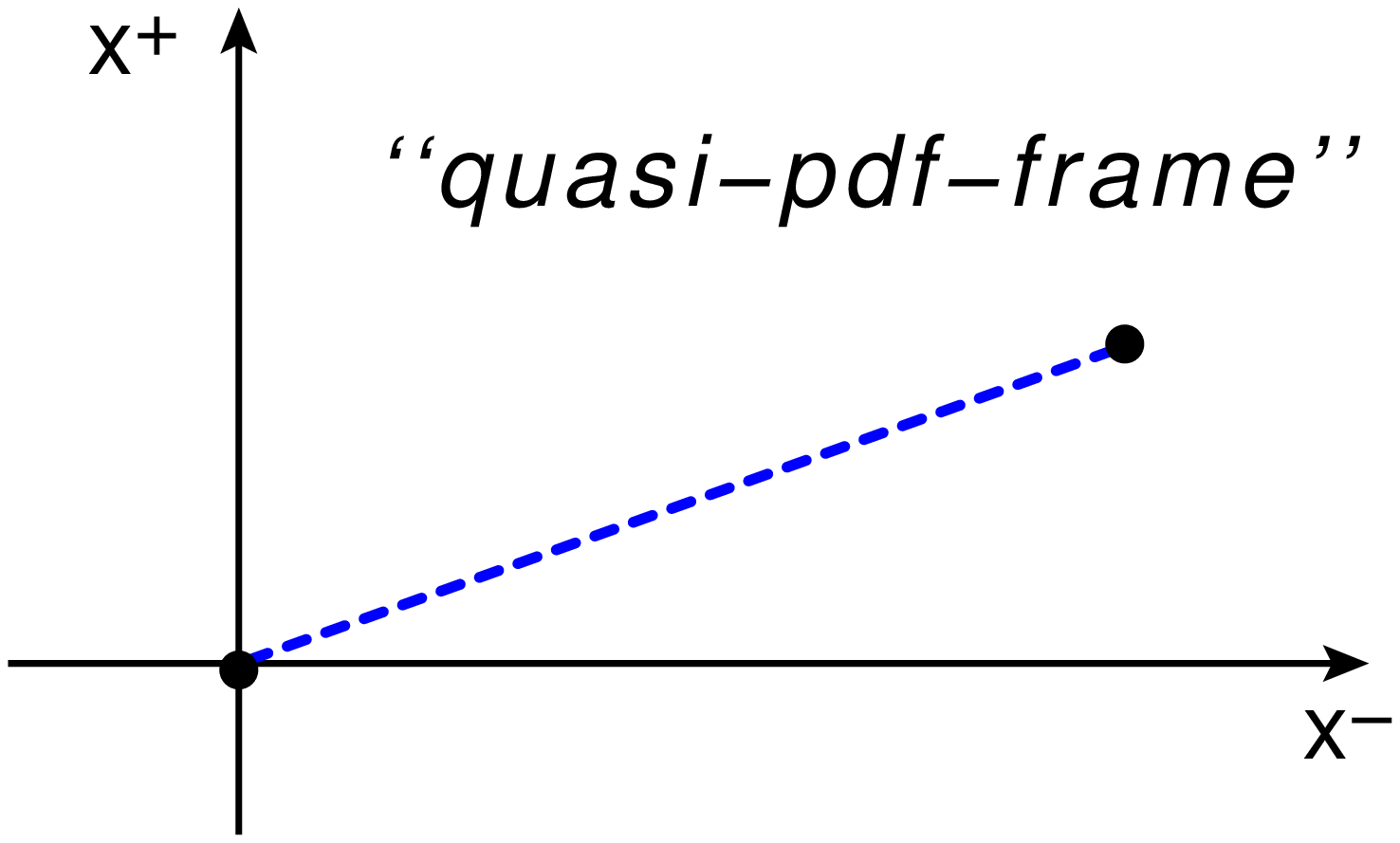}
\caption{In the left panel the Wilson frame is depicted, while in the right panel the quasi-pdf frame.}
\label{frames}
\end{center}
\end{figure}

The correlation functions with operators defined in eqs. (\ref{S1Forwardj})-(\ref{S3Forwardj})
in the BFKL limit, $j\to 1$, are divergent. 
A way to regulate such divergence is to consider the ``Wilson frame'' \cite{Balitsky:2013npa}
which are light-ray operators with point-splitting to regulate UV divergences (see Fig. \ref{frames} a))
\begin{eqnarray}
\hspace{-1cm}\calf^j_{p_1}(x_{1\perp},x_{2\perp})&\equiv& \int_0^{\infty}\!du\,u^{1-j}\calf_{p_1}(u; x_{1\perp},x_{2\perp})\,,\\
\hspace{-1cm}\calf_{p_1}(u;x_{1\perp},x_{2\perp})&\equiv& \int\!dv\, 2\Tr\Big\{F_{p_1\mu}(up_1+vp_1+x_{1\perp})[up_1+vp_1,vp_1]_{x_1}
\nonumber\\
&&\times[vp_1+x_{1\perp},vp_1+x_{2\perp}]{F_{p_1}}^\mu(vp_1+x_{2\perp})[vp_1,up_1+vp_1]_{x_2}
\Big\}\,.
\end{eqnarray}
In Ref. \cite{Balitsky:2013npa} the correlation function of two ``Wilson-frames'' 
\begin{eqnarray}
\langle \calf^j_{p_1}(x_{1\perp},x_{2\perp})\calf^{j'}_{p_2}(y_{1\perp},y_{2\perp})\rangle\,,
\end{eqnarray}
was calculated in the BFKL limit and the explicit expression for the structure function $C(\Delta,j)$ was derived.

The correlation function of ``Wilson-frames'' reminds us the 
correlation function of four $Z^2$ currents with $Z={1\over \sqrt{2}}(\phi_1+i\phi_2)$ a renorm-invariant chiral primary operator, 
in the BFKL limit \cite{Balitsky:2009yp} in $\cal N$=4 SYM theory, or the correlation function of four electromagnetic
currents in QCD which describes the $\gamma^*\gamma^*$ scattering in the Regge limit \cite{Chirilli:2014dcb}.

We will show that an alternative gauge-link geometry to regulate the UV divergences present in the BFKL limit
is the ``quasi-pdf frame'' (see Fig. \ref{frames} b)). Indeed, we will show that in the limit $\omega = j-1\to 0$, $g\to 0$, and 
$g^2\ll\omega\ll 1$ we get the same result as one obtained in Ref \cite{Balitsky:2013npa} with ``Wilson-frames''.

\subsection{Gluon correlation function with ``quasi-pdf frame''}
As announced in the previous section, we will calculate gluon correlation function with ``quasi-pdf frame'' in the BFKL limit.

The procedure is the same as the one adopted in Ref. \cite{Balitsky:2009yp, Balitsky:2018irv}, 
\textit{i.e.}, we will apply the high-energy OPE.

The gluon correlation function under consideration is
\begin{eqnarray}
\langle \calf^j_{p_1}(x_\perp,y_\perp)\calf^{j'}_{p_2}(x'_\perp,y'_\perp)\rangle\,,
\label{gluon-corre1}
\end{eqnarray}
with
\begin{eqnarray}
\calf^j_{p_1}(x_\perp,y_\perp) = \int_0^{+\infty}\!du^{1-j} \calf_{p_1}(u;x_\perp,y_\perp)\,,
\end{eqnarray}
and where now $\calf_{p_1}(x,y)$ is taken with quasi-pdf frame (see Fig. \ref{frames})
\begin{eqnarray}
\hspace{-0.7cm}\calf_{p_1}(u;x_\perp,y_\perp) = \!\!&&\int d v {F^{a}}_{p_1 i}(up_1+ vp_1+x_\perp)
\nonumber\\
&&\times[up_1+vp_1+x_\perp, vp_1+y_\perp]^{ab}{{F^b}_{p_1}}^i(vp_1+y_\perp)\,.
\label{quasipdf-def}
\end{eqnarray}
In coordinate space the Regge limit is achieved considering the limit
$x\!\cdot\! p_2, ~ x'\!\cdot\! p_1 \to \infty$, $y\!\cdot\! p_1, ~y'\!\cdot\! p_2\to -\infty$ 
and keeping all other components fixed. In this limit, the correlation function (\ref{gluon-corre1}) factorizes as~\cite{Balitsky:2009yp}
\begin{eqnarray}
\langle \calf_{p_1}(x_\perp,y_\perp)\rangle\langle\calf_{p_2}(x'_\perp,y'_\perp)\rangle\,.
\label{gluon-corre2}
\end{eqnarray}
To each factor of (\ref{gluon-corre2}) we apply the high-energy OPE similarly to what we did in section \ref{sec:saddapproxy}, 
indeed,  using result (\ref{dresult-diagram-differentiated}) we have
\begin{eqnarray}
&&\hspace{-1cm}
\int_0^{+\infty}\!dx^+\int_{-\infty}^0\!dy^+\delta(x^+-y^+-L)
\int_0^{+\infty}\!dx'^-\int_{-\infty}^0\!dy'^-\delta(x'^- - y'^- - L')
\nonumber\\
&&\hspace{-1cm}
\times
\langle F^{a\,i-}(x^+,x_\perp)[n x^+ + x_\perp, ny^+ + y_\perp]^{ab}F_i^{b\,-}(y^+,y_\perp)
\rangle_{{\rm Fig.}\ref{Fig:quasipdf-gq-LO}}
\nonumber\\
&&\hspace{-1cm}
\times
\langle F^{a'\,k+}(x'^-,x'_\perp)[n x'^- + x'_\perp, ny'^- + y'_\perp]^{a'b'}F_k^{b'\,+}(y'^-,y'_\perp)
\rangle_{{\rm Fig.}\ref{Fig:quasipdf-gq-LO}}
\nonumber\\
&&\hspace{-1cm}
=  {-6N^2_c\over (x^+|y^+|)^3}\int {d^2z_2\over \pi^3}\,\calu(z_1,z_2)
\Bigg[
{-2(x-z_2,y-z_2)^2_\perp + (x-z_2)^2_\perp(y-z_2)^2_\perp
	\over \left( {(y-z_2)^2_\perp\over |y^+|} + {(x-z_2)^2_\perp\over x^+}\right)^4}
\Bigg] 
\nonumber\\
&&\hspace{-1cm}
\times  {- 6N^2_c\over (x'^+|y'^+|)^3}\int {d^2z'_2\over \pi^3}\,\calu(z'_1,z'_2)
\Bigg[
{-2(x'-z'_2,y'-z'_2)^2_\perp + (x'-z'_2)^2_\perp(y'-z'_2)^2_\perp
\over \left( {(y'-z'_2)^2_\perp\over |y'^+|} + {(x'-z'_2)^2_\perp\over x'^+}\right)^4}
\Bigg]\,.
\label{gluon-corre3}
\end{eqnarray}
An important difference between the calculation of a correlation function like (\ref{gluon-corre1}),
and the calculation carried out in section \ref{sec:saddapproxy}, is that here we will not need a model to evaluate the
initial condition for the evolution equation of the matrix elements because the initial conditions are fully perturbative and
are obtained by calculating in pQCD the dipole-dipole scattering. In section \ref{sec:saddapproxy}, instead, we 
used a model to evaluate the dipole in the target state.

To proceed, we need the projection of the dipole-Wilson-line operator $\calu(z_1,z_2)$ onto the leading-order eigenfunctions, so 
using the completeness relation, we have
\begin{eqnarray}
&&\hspace{-2cm}\calu(z_1,z_2)= 
\int\!d^2z_0 \int {d\nu\over \pi^2}\,\nu^2\left({z^2_{12}\over z^2_{10}z^2_{20}}\right)^\gamma \calu(\nu,z_0)\,,
\label{calu-nu}
\end{eqnarray}
with $\gamma = \half + i\nu$ and $\bamma = 1-\gamma$, and where we defined 
\begin{eqnarray}
\calu(\nu,z_0) \equiv \int {d^2 z'_1 d^2z'_2\over\pi^2z'^4_{12}}
\left({z'^2_{12}\over z'^2_{10}z'^2_{20}}\right)^\bamma\calu(z'_1,z'_2)\,.
\end{eqnarray}
We also need the solution of the evolution equation for the dipole-Wilson-line operator in the linear case, \textit{i.e.} the BFKL equation,
\begin{eqnarray}
&&\calu^{Y_a}(\nu,z_0) = e^{(Y_a-Y_0)\aleph(\nu)}\calu^{Y_0}(\nu,z_0)\,,\\
&&\calu^{Y_b}(\nu,z'_0) = e^{(Y_0+Y_b)\aleph(\nu)}\calu^{Y_0}(\nu,z'_0)\,.
\end{eqnarray}
where $\aleph(\nu) = \aleph(\gamma(\nu))$ with $\gamma = \half + i\nu$.
The resummation parameter in coordinate space is \cite{Balitsky:2009yp}
\begin{eqnarray}
&& Y_a = \half\ln{2L^2\over (x-y)^2_\perp}\,, ~~~~~~~~ Y_b = \half\ln{2L'^2\over (x'-y')^2_\perp}\,.
\label{resumparameter-coord}
\end{eqnarray}
Indeed, in coordinate space, we already observed above, $L,L'\to \infty$ while the other components are kept fixed. 
To have an intuitive picture, one has to recall the more familiar case of $\gamma^*\gamma^*$ process, and think to
the $L$, and $L'$ as conjugated to the center of mass energy of the virtual photon-target system, and $(x-y)^2_\perp$,
and $(x'-y')^2_\perp$ conjugated to the virtuality of the photon.

So, using (\ref{calu-nu}), the dipole-dipole amplitude with BFKL resummation is (see section \ref {sec:dipodiposcat}
for details of the calculation)
\begin{eqnarray}
&&\hspace{-1cm}
\langle \calu^{Y_a}(z_1,z_2)\calu^{Y_b}(z'_1,z'_2)\rangle
\nonumber\\
&&\hspace{-1cm}
= - {\alpha^2_s (N^2_c-1)\over N^2_c}\int {d\nu \over \pi}{16\,\nu^2\over (1+4\nu^2)^2}
\Big({2LL'\over |(x-y)_\perp||(x'-y')_\perp|}\Big)^{\aleph(\gamma)}
\nonumber\\
&&\hspace{-1cm}
~~~\times{\Gamma^2(\half + i\nu)\Gamma(-2i\nu)\over\Gamma^2(\half - i\nu)\Gamma(1+2i\nu)} 
\left({z^2_{12}z'^2_{12}\over (X-X')^4}\right)^\gamma\,.
\label{dipo-dipo}
\end{eqnarray}

Using eq. (\ref{dipo-dipo}), and the result of projection eq. (\ref{projection}), from eq. (\ref{gluon-corre3}) we get
\begin{eqnarray}
&&\hspace{-1.2cm}
\int_0^{+\infty}\!dx^+\int_{-\infty}^0\!dy^+\delta(x^+-y^+-L)
\int_0^{+\infty}\!dx'^-\int_{-\infty}^0\!dy'^-\delta(x'^- - y'^- - L')
\nonumber\\
&&\hspace{-1.2cm}
~~\times\!
\langle F^{a\,i-}(x^+,x_\perp)[n x^+ + x_\perp, ny^+ + y_\perp]^{ab}F_i^{b\,-}(y^+,y_\perp)
\rangle_{{\rm Fig.}\ref{Fig:quasipdf-gq-LO}}
\nonumber\\
&&\hspace{-1.2cm}
~~\times\!
\langle F^{a'\,k+}(x'^-,x'_\perp)[n x'^- + x'_\perp, ny'^- + y'_\perp]^{a'b'}F_k^{b'\,+}(y'^-,y'_\perp)
\rangle_{{\rm Fig.}\ref{Fig:quasipdf-gq-LO}}
\nonumber\\
&&\hspace{-1.2cm}
=  {-9\alpha^2_s N^4_c \over LL' \Delta^2_\perp\Delta'^2_\perp}\int {d\nu \over \pi^5}
{64\,\nu^2\over (1+4\nu^2)^2}
{\gamma^{10}\Gamma(1-2\gamma)\Gamma^8(\gamma)\over \Gamma(2\gamma)\Gamma^2(2\gamma+2)}
{\left(\Delta^2_\perp\Delta'^2_\perp\over (X-X')_\perp^4\right)^\gamma}
\!\!\left({2L L'\over|\Delta^2_\perp||\Delta'^2_\perp|}\right)^{\!\aleph(\gamma)} \,,
\label{gluon-corre4}
\end{eqnarray}
where we defined $X_\perp = {x_\perp+y_\perp\over 2}$ and the same for $X'_\perp$, and 
where we used $N^2_c-1\to N^2_c$ in the large $N_c$ limit.
From (\ref{gluon-corre4}) we can calculate the correlation function of the $j$-dependent operators
\begin{eqnarray}
\hspace{-0.6cm}&&\langle \calf_{p_1}^j(x_\perp,y_\perp)\calf_{p_2}^{j'}(x'_\perp,y'_\perp)\rangle
\nonumber\\
&&=
-{9\alpha^2_s N^4_c \over LL' \Delta^2_\perp\Delta'^2_\perp}\left(s\over 2\right)^{j-1}\int_0^{+\infty}\!dL L^{1-j}\int_0^{+\infty}\!dL' L'^{1-j'}\theta(2LL' - (X-X')^2_\perp)
\nonumber\\
&&
~\times\!\!\int {d\nu \over \pi^5}
{64\,\nu^2\over (1+4\nu^2)^2}
{\gamma^{10}\Gamma(1-2\gamma)\Gamma^8(\gamma)\over \Gamma(2\gamma)\Gamma^2(2\gamma+2)}
{\left(\Delta^2_\perp\Delta'^2_\perp\over (X-X')_\perp^4\right)^\gamma}
\!\!\left({2L L'\over|\Delta^2_\perp||\Delta'^2_\perp|}\right)^{\!\aleph(\gamma)}\,,
\end{eqnarray}
where we used 
\begin{eqnarray}
\calf_{p_1}^j(x_\perp,y_\perp) = \!\!&&
\left(s\over 2\right)^{j-1\over 2}\int_0^{+\infty} dL \,L^{1-j}\int dx^+{F^{a-}}_\xi(x^++L,x_\perp)
\nonumber\\
&&\times[(x^++L)n+x_\perp, nx^++y_\perp]^{ab}F^{b\,-\xi}(x^+,y_\perp)\,,
\end{eqnarray}
and similarly for $\calf_{p_2}^j(x'_\perp,y'_\perp)$.
The $\theta(2LL' - (X-X')^2_\perp)$-function ensures that the longitudinal size of
two quasi-pdf frames are greater than the relative transverse separation.
Performing the integration over $L$ and $L'$ we arrive at
\begin{eqnarray}
&&\hspace{-1.2cm}\langle \calf^j_{p_1}(x_\perp,y_\perp)\calf^{j'}_{p_2}(x'_\perp,y'_\perp)\rangle
\nonumber\\
&&\hspace{-1.2cm}=
18{\alpha^2_sN^4_c\over \pi^4} \left(s\over 2\right)^{j-1}
\int d\nu\,
{2^\omega\,(\Delta^2_\perp\Delta'^2_\perp)^{\gamma-1-{\aleph(\gamma)\over 2}}
\over [(X-X')^2_\perp]^{2\gamma+j-1-\aleph(\gamma)}}\,
{\theta({\rm Re}[\omega - \aleph(\gamma)])\over \omega - \aleph(\gamma)}
\nonumber\\
&&~~\times
{(1-2\gamma)^2\over (1-\gamma)^2}
{\Gamma(1-2\gamma)\Gamma^8(1+\gamma)\over \Gamma(2\gamma)\Gamma^2(2\gamma+2)}\delta(\omega-\omega')\,.
\end{eqnarray}
To perform last integration we change variable $\nu \to \gamma = \half + i\nu$ and consider the ``DGLAP''
limit $\alpha_s\ll\omega=j-1\ll0$
\begin{eqnarray}
&&\hspace{-1.2cm}\langle \calf_{p_1}^j(x_\perp,y_\perp)\calf_{p_2}^{j'}(x'_\perp,y'_\perp)\rangle
\nonumber\\
&&\hspace{-1.2cm}=
- i\,18{\alpha^2_sN^4_c\over \pi^4} 
\int_{\half - i\infty}^{\half + i\infty}\!\!d\gamma\,
{s^\omega\,(\Delta^2_\perp\Delta'^2_\perp)^{\gamma-1-{\aleph(\gamma)\over 2}}
	\over [(X-X')^2_\perp]^{2\gamma+j-1-\aleph(\gamma)}}\,
{\theta({\rm Re}[\omega - \aleph(\gamma)])\over \omega - \aleph(\gamma)}
\nonumber\\
&&~~\times
{(1-2\gamma)^2\over (1-\gamma)^2}
{\Gamma(1-2\gamma)\Gamma^8(1+\gamma)\over \Gamma(2\gamma)\Gamma^2(2\gamma+2)}\delta(\omega-\omega')\,.
\label{gluon-corre5}
\end{eqnarray}

\begin{figure}
	\begin{center}
		\includegraphics[width=4in]{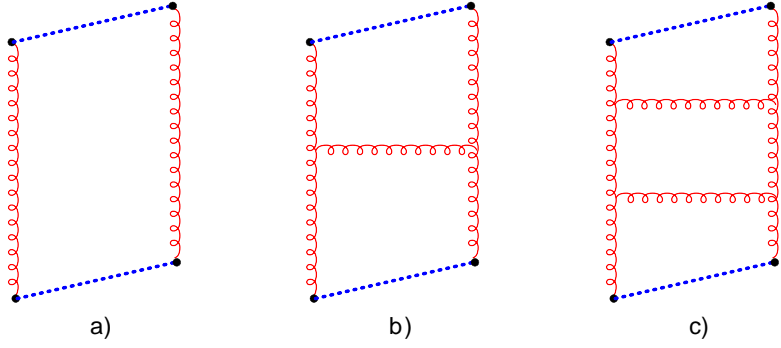}
	\end{center}
\caption{First three order diagrams for the correlator of two ``quasi-PDF'' frames. Diagrams a) and b) are not included in the
product of two dipole Wilson-line operators.}
\label{Fig:first3ordersDiagrams}
\end{figure}
In eq. (\ref{gluon-corre5}), we can close the contour to the right of 
all the residues and consider only the left most one which will give the
leading contribution. However, we observe that there are two residues to consider. The first, $\gamma = 1-{\alpha_s\over \omega}$,
reproduces the general expected result of eq. (\ref{2point-general}) in the high-energy limit. The second one, $\gamma=1$,
cancels out the diagrams a) and b) in Fig. \ref{Fig:first3ordersDiagrams} which are not included into the high-energy OPE formalism. 
To see this, one has to consider diagram in Fig. \ref{Fig:quasipdf-gq-LO} and expand the Wilson line operator to two gluon approximation. 
In this way, one observes that from this expansion, the first diagram that obtains is the one
in figure \ref{Fig:first3ordersDiagrams}c (plus permutations). Thus, diagrams in Fig. \ref{Fig:first3ordersDiagrams}a) and b)
are absent from the high-energy OPE, but they have to be included in the calculation of this correlation function. 
The contribution of diagram in figure \ref{Fig:first3ordersDiagrams} a) (diagram b) is higher order) will exactly 
cancel the contribution from the residue at the point $\gamma=1$. 
Diagram in Fig. \ref{Fig:first3ordersDiagrams} a) has been calculated in Ref.~\cite{Balitsky:2018irv}, so we do not need to
calculate it again, rather, we will show that it get canceled from the residue at $\gamma=1$
also in the case of quasi-pdf frame. We remind the reader that in Ref.~\cite{Balitsky:2018irv} (see also Ref.~\cite{Balitsky:2013npa})
the gluon correlation function was considered with Wilson frames (see Fig. \ref{frames}).

Taking the residue at $\aleph(\gamma^*) - \omega =0$ we have
\begin{eqnarray}
&&\hspace{-1.2cm}\langle \calf_{p_1}^j(x_\perp,y_\perp)\calf_{p_2}^{j'}(x'_\perp,y'_\perp)\rangle
\nonumber\\
&&\hspace{-1.2cm}=
36{\alpha^2_sN^4_c\over \pi^3} 
{s^\omega\,(\Delta^2_\perp\Delta'^2_\perp)^{\gamma^*-1-{\omega\over 2}}
\over [(X-X')^2_\perp]^{2\gamma^*}}\,
{(1-2\gamma^*)^2\over (1-\gamma^*)^2}
{\Gamma(1-2\gamma^*)\Gamma^8(1+\gamma^*)\over \Gamma(2\gamma^*)\Gamma^2(2\gamma^*+2)}
{\delta(\omega-\omega')\over \aleph'(\gamma^*)}\,.
\label{gluon-corre6}
\end{eqnarray}
Here $|\Delta_\perp||\Delta'_\perp|$ is the IR cut-off and $\aleph'(\gamma) = {d\over d\gamma}\aleph(\gamma)$.
Comparing (\ref{gluon-corre6}) with (\ref{2point-general}), we have two equations
$2\gamma^*-2-\omega=\gamma_{a}$ and $2\gamma^* = \Delta-1$, from which we get $\gamma_a+\omega = \Delta-3$
in agreement with result obtained in Ref.~\cite{Balitsky:2018irv}. 

In the limit $\alpha_s\ll\omega\ll1$ we have $\gamma^*=1-{\bar{\alpha}_s\over \omega}$ with 
$\bar{\alpha}_s = {\alpha_sN_c\over \pi}$, and result (\ref{gluon-corre6}) becomes
\begin{eqnarray}
\langle \calf_{p_1}^j(x_\perp,y_\perp)\calf_{p_2}^{j'}(x'_\perp,y'_\perp)\rangle
\simeq - N^2_c
{(|\Delta_\perp\Delta'_\perp|)^{-{2\bar{\alpha}_s\over\omega} - \omega}
\over [(X-X')^2_\perp]^{2-2{\bar{\alpha}_s\over \omega}}}
{\omega\over 2\pi}s^\omega\delta(j-j') \,.
\label{gluon-corre-res1}
\end{eqnarray}
Result (\ref{gluon-corre-res1}) coincides with the one obtained in Refs. \cite{Balitsky:2013npa, Balitsky:2018irv}
and is consistent with the general result for two-point correlation function eq. (\ref{2point-general}).

What we are left to do is the calculation of the residue at $\gamma=1$ and show that it coincides with the one calculated in
\cite{Balitsky:2018irv}, thus canceling the contribution of diagrams  a) and b) in figure \ref{Fig:first3ordersDiagrams}.
So, we start from eq. (\ref{gluon-corre5}) which we can rewrite as
\begin{eqnarray}
&&\hspace{-1.2cm}\langle \calf_{p_1}^j(x_\perp,y_\perp)\calf_{p_2}^{j'}(x'_\perp,y'_\perp)\rangle
\nonumber\\
&&\hspace{-1.2cm}=
- i\,18{\alpha^2_sN^4_c\over \pi^4} 
\int_{\half - i\infty}^{\half + i\infty}\!\!d\gamma\,
{s^\omega\,(\Delta^2_\perp\Delta'^2_\perp)^{\gamma-1-{\aleph(\gamma)\over 2}}
	\over [(X-X')^2_\perp]^{2\gamma+j-1-\aleph(\gamma)}}\,
{\theta({\rm Re}[\omega - \aleph(\gamma)])\over \omega - \aleph(\gamma)}
\nonumber\\
&&~~\times
{(1-2\gamma)^2\over 2(1-\gamma)^3}
{\Gamma(3-2\gamma)\Gamma^8(1+\gamma)\over \Gamma(2\gamma)\Gamma^2(2\gamma+2)}\delta(j-j')\,.
\label{gluon-corre7}
\end{eqnarray}
The residue at $\gamma = 1$ is
\begin{eqnarray}
\hspace{-0.7cm}\langle \calf_{p_1}^j(x_\perp,y_\perp)\calf_{p_2}^{j'}(x'_\perp,y'_\perp)\rangle
= \!\!&& - {N^2_c s^\omega  \omega\over \pi[(X-X')^2_\perp]^{\omega+2}}
\left( {\bar{\alpha}_s\over 3\omega}
- \half - {\bar{\alpha}_s\over \omega}\ln{(X-X')^2_\perp\over |\Delta_\perp||\Delta'_\perp|}\right)
\delta(j-j')
\nonumber\\
\simeq\!\!&& {N^2_c  s^\omega \omega\over 2\pi[(X-X')^2_\perp]^{\omega+2}}
\left(1 + 2{\bar{\alpha}_s\over \omega}\ln{(X-X')^2_\perp\over |\Delta_\perp||\Delta'_\perp|}\right)
\delta(j-j')\,,
\label{secondpole}
\end{eqnarray}
where we used $\bar{\alpha}_s\ll\omega\ll 1$ and $\bar{\alpha}_s\ll\bar{\alpha}_s\ln{(X-X')^2_\perp\over |\Delta_\perp||\Delta'_\perp|}$.
Result (\ref{secondpole}) coincides exactly with eq. (5.40) of reference \cite{Balitsky:2018irv}.

In conclusion, in this section we have proven that the Wilson-frame regulator
and the quasi-pdf frame regulator give the same result in the calculation of the two-point correlation function in the high-energy, $j\to 1$, limit.
This justifies the use of the HE-OPE for the calculation of the high-energy behavior of the LT and NLT gluon distributions.

\section{Dipole-dipole scattering}
\label{sec:dipodiposcat}

In this section we calculate the dipole-dipole scattering.
Using (\ref{calu-nu}), we have
\begin{eqnarray}
	&&\hspace{-1cm}
	\langle \calu^{Y_a}(z_1,z_2)\calu^{Y_b}(z'_1,z'_2)\rangle
	\nonumber\\
	&&\hspace{-1cm}
	= \int {d\nu\over \pi^2}\,\nu^2\!\int d^2 z_0\left({z^2_{12}\over z_{10}^2z_{20}^2}\right)^\gamma
	\!\!\int\!\! {d\nu'\over \pi^2}\,\nu'^2\!\int\!\! d^2 z'_0\left({z'^2_{12}\over z'^2_{10}z'^2_{20}}\right)^{\gamma'}
	\!\!\langle \calu^{Y_a}(z_0,\nu)\calu^{Y_b}(z'_0,\nu')\rangle\,,
	\label{dipo-dipo1}
\end{eqnarray}
where, as we explained above, the evolution parameters are $Y_a = \half\ln {2L^2\over (x-y)^2_\perp}$ and 
$Y_b = \half\ln {2L'^2\over (x'-y')^2_\perp}$. Assuming that $Y_0$ is the initial condition for the evolution,
the dipole-dipole scattering with BFKL resummation is
\begin{eqnarray}
\hspace{-1cm}\langle\calu^{Y_a}(z_0,\nu)\calu^{Y_b}(z_0,\nu)\rangle 
=\!\!\!&& \Big\langle e^{(Y_a-Y_0)\aleph(\nu)}\calu^{Y_0}(\nu,z_0)e^{(Y_0+Y_b)\aleph(\nu)}\calu^{Y_0}(\nu,z_0)
\Big\rangle
\nonumber\\
&&\times e^{(Y_a+Y_b)\aleph(\nu)}
{-\alpha_s^2(N_c^2-1)\over 4 N_c^2}
{16\pi^2\over \nu^2(1+4\nu^2)^2}
\nonumber\\
&&\times\Big[\delta(z_0-z'_0)\delta(\nu+\nu')
+{2^{1-4i\nu}\delta(\nu-\nu')\over \pi|z_0-z'_0|^{2-4i\nu}}
{\Gamma\big({1\over 2}+i\nu\big)\Gamma(1-i\nu)\over\Gamma(i\nu)\Gamma\big(\half-i\nu\big)}\Big]\,.
\label{evolution-dipo-ab}
\end{eqnarray}
Substituting (\ref{evolution-dipo-ab}) in eq. (\ref{dipo-dipo1}) we have
\begin{eqnarray}
	\hspace{-1cm}
	\langle\calu^{Y_a}(z_0,\nu)\calu^{Y_b}(z_0,\nu)\rangle 
	=\!\!\!&& -{\alpha^2_s (N^2_c-1)\over N^2_c}\int {d\nu \over \pi}{16\,\nu^2\over (1+4\nu^2)^2}
	\Big({2LL'\over |(x-y)_\perp||(x'-y')_\perp|}\Big)^{\aleph(\gamma)}
	\nonumber\\
	&&\times{\Gamma^2(\half + i\nu)\Gamma(-2i\nu)\over\Gamma^2(\half - i\nu)\Gamma(1+2i\nu)} 
	\left({z^2_{12}z'^2_{12}\over (X-X')^4}\right)^\gamma\,,
	\label{dipo-dipo2}
\end{eqnarray}
with $X={x_\perp+y_\perp\over 2}$ and the same for $X'$, and where we used
\begin{eqnarray}
&&\int d^2 z_0\left({z^2_{12}\over z_{10}^2z_{20}^2}\right)^\gamma
	\left({z'^2_{12}\over (z'_1 - z_0)^2(z'_2-z_0)^2}\right)^{1-\gamma}
	\nonumber\\
	&&
= \left\{\left({z^2_{12}z'^2_{12}\over (X-X')^4}\right)^\gamma
	\pi{\Gamma^2(\half + i\nu)\Gamma(-2i\nu)\over\Gamma^2(\half - i\nu)\Gamma(1+2i\nu)} + \nu\leftrightarrow -\nu\right\}\,,
	\label{firstterm}
\end{eqnarray}
and
\begin{eqnarray}
	&&\hspace{-2cm}\int {d\nu d\nu'\over \pi^4}\nu^2\nu'^2\int d^2 z_0\left({z^2_{12}\over z_{10}^2z_{20}^2}\right)^\gamma
	\int d^2 z'_0\left({z'^2_{12}\over z'^2_{10}z'^2_{20}}\right)^{\gamma'}
	\nonumber\\
	&&\hspace{-2cm}\times
	{2^{1-4i\nu}\delta(\nu-\nu')\over \pi|z_0-z'_0|^{2-4i\nu}}{\Gamma(\half+i\nu)\Gamma(1-i\nu)\over \Gamma(i\nu)\Gamma(\half-i\nu)}
	\nonumber\\
	&&\hspace{-2cm}
	= \int {d\nu \over \pi^4}\nu^4\left\{
	\left({z^2_{12}z'^2_{12}\over (X-X')^4}\right)^\gamma
	\pi{\Gamma^2(\half + i\nu)\Gamma(-2i\nu)\over\Gamma^2(\half - i\nu)\Gamma(1+2i\nu)} + \nu\leftrightarrow -\nu\right\}\,.
	\label{secondterm}
\end{eqnarray}

\section{LT and NLT pseudo-PDF: analytic expression}
\label{sec:NLTanalytic-pseudoPDF}

\begin{figure}
	\begin{center}
		\includegraphics[width=4in]{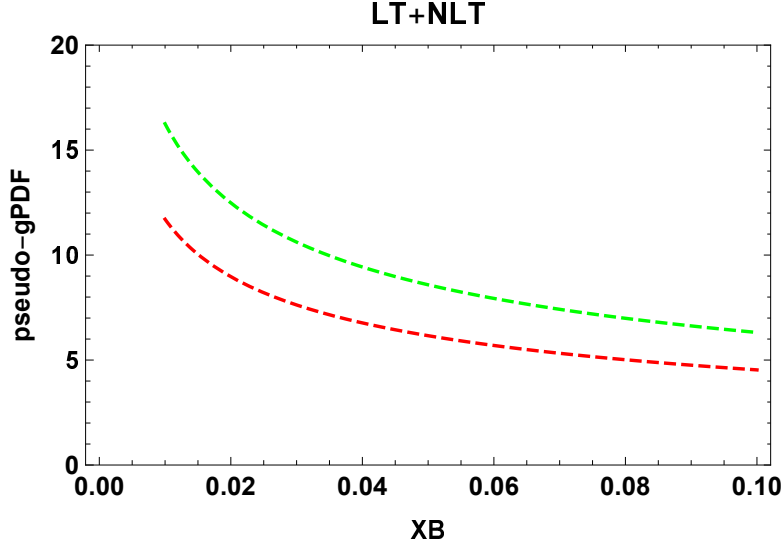}
	\end{center}
	\caption{Comparing the numerical evaluation of eq. (\ref{fourier-inversmellin-g}) (red dashed curve) with its
		approximated analytic result, eq. (\ref{fourier-inversmellin-g-approxyanaly}) (green dashed curve).}
	\label{Fig:pseudoNLTtwisAnalytic}
\end{figure}

In this section we evaluate the inverse Mellin transform of eq. (\ref{fourier-inversmellin-g-approxy}) which is the approximated
result of eq. (\ref{fourier-inversmellin-g}).
To perform the inverse Mellin, regardless of whether $\ln {4\over Q_s^2|z|^2}$ is positive or negative, is 
\begin{eqnarray}
	\hspace{-1.cm}
	G_{\rm p}(x_B,z^2)
	=\!\!\!&&
	{N^2_c\,Q^2_s\sigma_0\over 16\bar{\alpha}_s\pi^3}
	{1\over 2\pi i} \int_{1-i\infty}^{1+i\infty} {d\omega\over \omega}\,
	\left({2\over x_B^2|z|^2M_N^2}\right)^{\omega\over 2}
	\left({4\over Q^2_s|z|^2}\right)^{{\bar{\alpha}_s\over \omega}} 	
	\nonumber\\
	\hspace{-1.cm}
	&&\times\left(1 +{Q_s^2|z|^2\over 5}\right) + O\!\left({Q_s^4|z|^4\over 16}\right)
	\nonumber\\
	=\!\!\!&&{N^2_c Q_s^2\sigma_0\over 16 \pi^3\bar{\alpha}_s}\left(1 +{Q_s^2|z|^2\over 5}\right) 
	I_0(h)
	+ O\!\left({Q_s^4|z|^4\over 16}\right)\,,
	\label{fourier-inversmellin-g-approxyanaly}
\end{eqnarray}
with $h$ defined as
\begin{eqnarray}
	h = \left[2\bar{\alpha}_s\left|\ln{4\over |z|^2Q_s^2}\right|\ln{2\over x_B^2|z^2|M^2_N}\right]^\half\,.
\end{eqnarray}
In Fig. \ref{Fig:pseudoNLTtwisAnalytic} we show that eq. (\ref{fourier-inversmellin-g}) can be well approximated by
result (\ref{fourier-inversmellin-g-approxyanaly}). 

\section{LT and NLT quasi-PDF: analytic expression}
\label{sec:analytic-qPDFtwist}

To perform the inverse Mellin transform in eq. (\ref{quasiPDFtwistapproxy}), we have to distinguish two cases. 
The first case is when 
$\ln{4P_\xi^2x_B^2\over Q^2_s}<0$, and the inverse Mellin transform is 
\begin{eqnarray}
G_{\rm q}(x_B,P_\xi)&&
\simeq {N^2_c\,Q^2_s\,\sigma_0\over 16\bar{\alpha}_s\pi^3} \left[
{\ln\left(-{Q_s^2\over 4 P_\xi^2 x_B^2}-i\epsilon\right)\over \ln\left(-{2P_\xi^2\over M^2_N}+i\epsilon\right)}
\left(J_0(m) - J_2(m) - {2\over m}J_1(m)\right)\right.
\nonumber\\
&&\left. ~~~~~~~~~~~~~~~~~ + {2Q_s^2\over 5 P_\xi^2 x_B^2}
\left({2\bar{\alpha}_s\ln\left(-{Q_s^2\over4 P_\xi^2 x_B^2} - i\epsilon \right)
\over \ln\left(-{2 P_\xi^2\over M_N^2}+i\epsilon\right)}\right)^\half J_1(m)
\right]\,,
\label{quasiPDFtwistapproxybb}
\end{eqnarray}
where we defined 
\begin{eqnarray}
m\equiv  \left[2\bar{\alpha}_s\ln\left(-{Q_s^2\over4 P_\xi^2 x_B^2} - i\epsilon \right)\ln\left(-{2 P_\xi^2\over M_N^2}+i\epsilon\right)
\right]^\half\,.
\end{eqnarray}

\begin{figure}[htb]
	\begin{center}
		\includegraphics[width=3.8in]{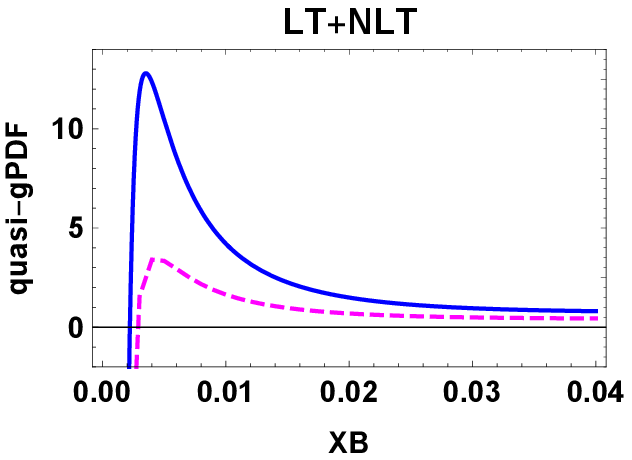}
		\caption{We compare the real part of eq. (\ref{quasiPDFtwist}) (magenta-dashed curve) with its approximation, 
			real part of eq. (\ref{quasiPDFtwistapproxybb}) (blue curve).
			This is the case $\ln{4P_\xi^2x_B^2\over Q^2_s}<0$, with $P_\xi=4$\, GeV, and $x_B\in [0.001, 0.04]$.}
		\label{Fig:quasiPDFtwistapproxy}
	\end{center}
\end{figure}

The second case is when $\ln{4P_\xi^2x_B^2\over Q^2_s}>0$, and the inverse Mellin transform is 
\begin{eqnarray}
\hspace{-2cm}G_q(x_B, P_\xi) && \!\!\simeq
- {N^2_c Q_s^2 \sigma_0\over 16\bar{\alpha}_s\pi^3}
\left[
{\ln\left(-{4P_\xi^2 x_B^2\over Q_s^2}+i\epsilon\right)
\over \ln\left(-{2P_\xi^2\over M^2_N}+i\epsilon\right)}\left(I_0(\tilde{m}) + I_2(\tilde{m}) - {2\over \tilde{m}}I_1(\tilde{m}) \right)
\right.
\nonumber\\
&&
\left. ~~~ + {2 Q_s^2\over 5 P^2_\xi x_B^2}\left({2\bar{\alpha}_s}\ln\left(-{4P_\xi^2 x_B^2\over Q_s^2}+i\epsilon\right)
\over \ln\left(-{2P_\xi^2\over M^2_N}+i\epsilon\right)\right)^\half I_1(\tilde{m})\right]\,,
\label{quasiPDFtwistapproxycc}
\end{eqnarray}
and where we defined 
\begin{eqnarray}
\tilde{m}
\equiv  \left[2\bar{\alpha}_s\ln\left(-{4P_\xi^2 x_B^2\over Q_s^2}+i\epsilon \right)\ln\left(-{2 P_\xi^2\over M_N^2}+i\epsilon\right)
\right]^\half\,.
\end{eqnarray}

\begin{figure}[htb]
	\begin{center}
		\includegraphics[width=3.8in]{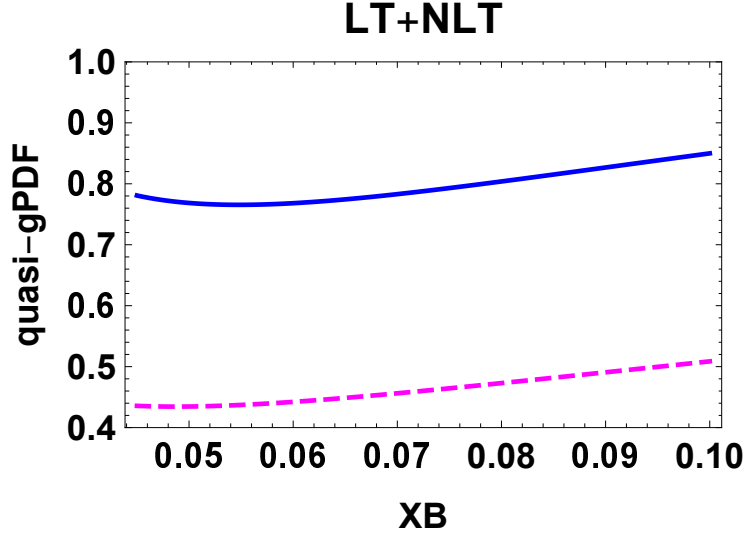}
		\caption{Here, we compare the real part of eq. (\ref{quasiPDFtwist}) (magenta-dashed curve) 
			with its approximation, real part of eq. (\ref{quasiPDFtwistapproxycc}) (blue curve).
		This time we are in the case $\ln{4P_\xi^2x_B^2\over Q^2_s}>0$, with $P_\xi=4$\,GeV, and $x_B\in [0.045, 0.1]$.}
		\label{Fig:quasiPDFtwistapproxy2}
	\end{center}
\end{figure}

\bibliographystyle{JHEP}
\bibliography{/Users/chirilli/Documents/mm/m/MyReferences}

\end{document}